\documentclass[acmsmall,screen]{acmart}

\usepackage{enumerate}

\usepackage{subfigure}
\usepackage{multirow}

\usepackage[most]{tcolorbox}
\usepackage{threeparttable}

\usepackage{colortbl}
\usepackage[table,xcdraw]{xcolor}
\usepackage{lipsum} 

\usepackage{circledsteps}

\usepackage{adjustbox}

\definecolor{Purple1}{rgb}{0.55, 0.3, 0.7} 
\definecolor{Purple2}{rgb}{0.6, 0.4, 0.7}
\definecolor{Purple3}{rgb}{0.7, 0.5, 0.8}
\definecolor{Purple4}{rgb}{0.8, 0.6, 0.9}
\definecolor{Purple5}{rgb}{0.9, 0.8, 1.0} 
\definecolor{Purple6}{rgb}{0.95, 0.85, 1.0} 
\definecolor{Purple7}{rgb}{0.98, 0.92, 1.0} 

\lstdefinestyle{javastyle}{
  language=Java,
  basicstyle=\ttfamily\tiny,
  numbers=left,
  numberstyle=\tiny\color{gray},
  stepnumber=1,
  numbersep=5pt,
  tabsize=4,
  showspaces=false,
  showstringspaces=false,
  breaklines=true,
  captionpos=b
}

\AtBeginDocument{%
  }

\setcopyright{acmlicensed}
\copyrightyear{2025}
\acmYear{2025}
\acmDOI{XXXXXXX.XXXXXXX}

\begin{document}

\title{WITNESS: A lightweight and practical approach to fine-grained predictive mutation testing}

\author{Zeyu Lu}
\affiliation{%
  \institution{State Key Laboratory for Novel Software Technology, Nanjing University}
  \city{Nanjing}
  \country{China}}
\email{zeyulu@smail.nju.edu.cn}

\author{Peng Zhang}
\affiliation{%
  \institution{University of Luxembourg}
  \city{Luxembourg}
  \country{Luxembourg}
}
\email{peng.zhang@uni.lu}

\author{Chun Yong Chong}
\affiliation{%
  \institution{Huawei Technologies Co., Ltd.}
  \city{Hong Kong}
  \country{China}
}
\email{chong.chun.yong@huawei.com}

\author{Shan Gao}
\affiliation{%
  \institution{Independent Researcher}
  \city{Hang Zhou}
  \country{China}
  }
\email{gaoshan_cs@outlook.com}

\author{Yibiao Yang, Yanhui Li, Lin Chen, Yuming Zhou}
\affiliation{%
  \institution{State Key Laboratory for Novel Software Technology, Nanjing University}
  \city{Nanjing}
  \country{China}}
\email{{yangyibiao, yanhuili, lchen}@nju.edu.cn, zhouyuming@nju.edu.cn}

\renewcommand{\shortauthors}{Lu et al.}

\begin{abstract}
\textbf{Background:} Predictive mutation testing mitigates the high computational cost of traditional mutation testing by employing prediction models to approximate mutation testing results. In its fine-grained form, predictive mutation testing predicts whether individual test cases will kill specific mutants. These predictions enable the construction of kill matrices, which is essential for downstream tasks such as test case prioritization. \textbf{Problem:} Existing fine-grained predictive mutation testing studies predominantly rely on deep learning, which faces two critical limitations in practice: (1) Exorbitant computational costs. The deep learning models adopted in these studies demand significant computational resources for training and inference acceleration. This introduces high costs and undermines the cost-reduction goal of predictive mutation testing. (2) Constrained applicability. Although modern mutation testing tools generate mutants both inside and outside methods, current fine-grained predictive mutation testing approaches handle only inside-method mutants. As a result, they cannot predict outside-method mutants, limiting their applicability in real-world scenarios. \textbf{Objective:} We aim to develop a more cost-effective and practically applicable fine-grained predictive mutation testing approach that reduces computational costs while supporting the prediction of outside-method mutants, thereby enhancing its real-world applicability. \textbf{Method:} We propose WITNESS, a new fine-grained predictive mutation testing approach. WITNESS adopts a twofold design: (1) With collected features from both inside-method and outside-method mutants, WITNESS is suitable for all generated mutants. (2) Instead of using computationally expensive deep learning, WITNESS employs lightweight classical machine learning models for training and prediction. This makes it more cost-effective and enabling straightforward explanations of the decision-making processes behind the adopted models. \textbf{Result:} Evaluations on Defects4J projects show that WITNESS consistently achieves state-of-the-art predictive performance across different scenarios. Additionally, WITNESS significantly enhances the efficiency of kill matrix prediction. Post-hoc analysis reveals that features incorporating information from before and after the mutation are the most important among those used in WITNESS. Test case prioritization based on the predicted kill matrix shows that WITNESS delivers results much closer to those obtained by using the actual kill matrix, outperforming baseline approaches. \textbf{Conclusion:} Compared to deep learning-based fine-grained predictive mutation testing, WITNESS is simpler yet delivers higher effectiveness, greater efficiency, and broader applicability. It advances fine-grained predictive mutation testing and sets a new baseline for future research.
\end{abstract}

\ccsdesc[500]{Software and its engineering~Software testing and debugging}

\keywords{Mutation Testing, Machine Learning, Kill Matrix, Test Case Prioritization}


\maketitle

\section{Introduction}

Mutation testing, which was first proposed by DeMillo \cite{demillo1978hints}, is widely used to evaluate the effectiveness of test suites. Mutation testing employs mutation operators to modify the original program under test, creating many mutants. These modifications are small, making the mutants identical to the original program except for the changed statements. The test suite is executed against each mutant to detect differences in output compared to the original program. If a test case causes a mutant to produce different outputs, the mutant is considered ‘‘killed.’’ Conversely, if no test case can produce different outputs, the mutant ‘‘survives.’’ The effectiveness of the test suite is measured by the mutation score, which is the ratio of killed mutants to the total number of mutants. A higher mutation score indicates a more effective test suite, while a lower score suggests weaker defect-detection capability.

Although mutation testing is more effective than code coverage in evaluating test suite effectiveness \cite{wong1995fault, offutt1996experimental, frankl1997all, baker2012empirical, just2014mutants}, it is less commonly adopted in practice \cite{sanchez2024mutation}. The primary reason for this is the high cost associated with mutation testing. Without optimization, the test suite must be executed on each mutant to calculate the mutation score. Even a small program could generate many mutants \cite{zhang2010operator}, and executing the test suite against all generated mutants requires substantial time, especially for larger projects. To reduce the cost of mutation testing, many approaches have been proposed. Mutant reduction \cite{guizzo2020sentinel, zhang2013operator} is one example. By decreasing the number of executed mutants, the cost of mutation testing can be reduced. However, even with mutant reduction, the execution of the test suite against some generated mutants is still required, keeping the costs relatively high \cite{mao2019extensive}. 

Predictive mutation testing uses machine learning models to predict whether a mutant will be killed, offering a more cost-effective alternative to traditional mutation testing. This approach was first introduced by Zhang et al. \cite{zhang2018predictive}, who used a Random Forest model to predict whether a mutant will be killed by the entire test suite based on specific features. However, predicting mutant killing outcomes does not fully capture the relationship between individual mutants and test cases. To tackle this problem, three recent studies proposed Seshat \cite{kim2022predictive}, MutationBERT \cite{jain2023contextual}, and SODA \cite{zhao2024spotting}, to predict whether a specific test case can kill a given mutant, generating a predicted kill matrix for all mutant-test case pairs. This finer-grained approach allows for more detailed downstream tasks, such as mutation-based test case prioritization using the Additional approach \cite{shin2019empirical}. 

Despite these advancements, Seshat, MutationBERT, and SODA all have notable limitations. 

\Circled{1} \textbf{High computational cost.} These approaches use deep learning models that entail high computational demands for efficiency. Such demands introduce higher costs, reduced accessibility, and more complex setup and maintenance, contradicting the cost-reduction goal of predictive mutation testing. For example, the total computational expense we incurred for conducting experiments across all three studies \cite{kim2022predictive, jain2023contextual, zhao2024spotting} in this paper exceeded \$3,000. Furthermore, even with the aid of computational resources, the prediction process remains time-consuming; for instance, we found that MutationBERT’s prediction time exceeded the time required for actual mutation testing in several experiments on our machine. 
 
\Circled{2} \textbf{Constrained applicability.} These approaches are applicable only to mutants where the mutation occurs within source methods. As they rely on the names and parameters of source methods, MutationBERT and SODA also collect contextual statements surrounding a mutated statement within the same source method. Outside-method statements—whose types typically include field declarations, static initializer blocks, instance initializer blocks, and definitions of inner classes, enums, or interfaces—are essential for defining the structure, state, and initialization behavior of Java classes \cite{alves1999formal}. Existing mutation testing tools \cite{just2014major, coles2016pit, schuler2009javalanche, ma2006mujava} for Java do not restrict their applicability to inside-method statements alone. For example, for mutants generated by Major \cite{just2014major}, a mutation testing tool commonly used by Seshat, MutationBERT, and SODA, the mutant–test pairs with mutations occurring outside source methods can account for up to 24.35\% of all mutant–test pairs in the JacksonCore\_5 project used in our study. This significantly limits the utility of these approaches.

To address the limitations of the three approaches that predict the kill matrix, we propose WITNESS (light-\underline{W}e\underline{I}ght \underline{T}est effective\underline{NES}s mea\underline{S}urement), a lightweight approach that evaluates test suite effectiveness by predicting the kill matrix. WITNESS adopts a dual-track design: (1) With collected features tailored to both inside-method and outside-method mutants, WITNESS is suitable for all generated mutants, thereby facilitating its practical use. (2) WITNESS employs lightweight classical machine learning models (Random Forest and LightGBM) for training and prediction. It is simpler, more cost-effective, and offers better accessibility compared to computationally expensive deep learning approaches. Moreover, deep learning models are often "black boxes" with limited interpretability, making it difficult to understand why a specific test case kills a particular mutant and complicating the explanation of the decision-making process. Although these machine learning models are non-linear like deep learning models, post-hoc analysis is much easier to perform on them, enabling a better understanding of their decision-making processes.

WITNESS begins by collecting features from three sources: the source code, the changes made to the source code, and the test cases. For each mutant-test pair, WITNESS employs machine learning models to predict whether the test case will kill the mutant. We evaluated Seshat, MutationBERT, SODA, and WITNESS across same-version, cross-version, and cross-project scenarios. In the same-version scenario, training and prediction are conducted within a single version of a project. In the cross-version scenario, mutation testing results from earlier versions are used to predict outcomes in later versions of the same project. In the cross-project scenario, source projects are utilized to predict the mutation testing results of target projects. Our experimental results indicate that WITNESS generally outperforms Seshat, MutationBERT, and SODA, demonstrating higher predictive performance across various killing reasons. WITNESS also achieves better results than the baselines in predicting fewer-covered mutants (i.e., mutants covered by fewer test cases among all test cases that could cover them). Our findings show that features incorporating information from before and after the mutation are the most influential in WITNESS. As a lightweight approach, WITNESS is significantly more efficient than Seshat, MutationBERT, and SODA. Using the predicted kill matrix, we conducted mutation-based test case prioritization. The results indicate that WITNESS’s predicted kill matrix yields prioritization results that are closer to those obtained using the actual kill matrix than those from the predicted matrices of the three baselines. WITNESS demonstrates that classical machine learning can achieve superior effectiveness, broader applicability, and drastically higher efficiency compared to deep learning-based fine-grained predictive mutation testing. It establishes a stronger, more practical baseline for fine-grained predictive mutation testing and advances the field toward real-world application, demonstrating that simpler methods can indeed be more effective.

Overall, our paper makes the following contributions:

\begin{enumerate}[(1)]
\item We propose a fine-grained predictive mutation testing approach called WITNESS, which uses machine learning to efficiently predict the kill matrix. WITNESS effectively predicts the kill matrix for all generated mutants, overcoming the limitations of baseline approaches, which are only suitable for mutants where mutations occur within source methods. 

\item	We conduct a thorough evaluation (involving 128 experiments) of fine-grained predictive mutation testing approaches, using significantly more experiments across different scenarios than the baselines. This overcomes the limitation of existing studies that conducted experiments under a single scenario or with only a limited number of experiments (e.g., only two experiments in MutationBERT).

\item We demonstrate that WITNESS generally achieves higher predictive performance than the baselines. WITNESS eliminates the high computational costs associated with GPU-based training and prediction. On average, WITNESS accelerates prediction times by 65.92 times, 1,722.81 times, and 986.17 times compared to Seshat, MutationBERT, and SODA, respectively. Although WITNESS is a lightweight approach, it remains simple yet effective.

\item We analyze the importance of different features used by WITNESS to understand the decision-making process employed by the machine learning models. This is the first study to analyze feature importance in fine-grained predictive mutation testing studies.

\end{enumerate}

The rest of the paper is organized as follows. Section 2 provides an overview of related work. Section 3 illustrates the proposed approach for predicting the kill matrix. Section 4 details the experimental setup. Section 5 presents the comparison results of the proposed approach and the baselines. Section 6 presents further experimental results and discusses the implications of the study and the threats to validity. Section 7 concludes the paper.

\section{Background and Related Work}

This section presents background and related work on reducing the cost of mutation testing and provides a detailed comparison of existing predictive mutation testing approaches.

\subsection{Mutation Cost Reduction}

Offutt and Untch \cite{offutt2001mutation} classified three strategies to reduce the cost of mutation testing: ‘do fewer’, ‘do smarter’, and ‘do faster’. The ‘do fewer’ strategies involve generating and executing fewer mutants. Examples include selective mutation (operator) \cite{namin2008sufficient}, mutant sampling \cite{gopinath2015hard}, higher order mutation \cite{harman2014angels}, mutant clustering \cite{hussain2008mutation}, and subsuming mutants \cite{souza2020identifying, ojdanic2022use}. The ‘do smarter’ strategies aim to distribute the execution of mutants across multiple computers or avoid full execution of mutants. For instance, in weak mutation \cite{androutsopoulos2014analysis}, the execution stops once the mutated statement is executed. The ‘do faster’ strategies focus on generating and executing mutants more quickly, for example, by using compiler-associated optimizations. Mutant schemata \cite{untch1997tums} fall into this category.

Recently, predictive mutation testing has emerged as another way to reduce the cost of mutation testing. Zhang et al. \cite{zhang2018predictive} proposed the first predictive mutation testing approach (PMT), which predicts mutant execution results without actual execution. They identified 14 features across three groups based on PIE (Propagation, Infection, and Execution) theory. Specifically, the three groups of features are: execution features, such as \textit{numExecuteCovered}; infection features, such as \textit{typeOperator}; and propagation features, such as \textit{depInheritance}. The experimental results show that the proposed approach performs well in both cross-version and cross-project scenarios. Later, Mao et al. \cite{mao2019extensive} extended the feature set of Zhang et al. \cite{zhang2018predictive} to include 95 features and adopted 11 classification models. They identified 91 static features, categorized into method metrics (e.g., \textit{classesReferenced}), class metrics (e.g., \textit{fanIn}), and package metrics (e.g., \textit{numberOfClasses}). The four dynamic features are \textit{numExecuted}, \textit{numTestCover}, \textit{numAssertInTM}, and \textit{numAssertInTC}. They found that predictive mutation testing performs well in cross-project scenarios. Zhang et al. \cite{zhang2020cbua} proposed CBUA, an unsupervised predictive mutation testing approach. It predicts the probability that a mutant survives, then calculates the number of survived mutants to further obtain the predicted mutation score. 

However, Aghamohammadi et al. \cite{aghamohammadi2021ensemble} highlight that the impact of unreached mutants significantly affects the predictive performance of predictive mutation testing. Their results indicate that predictive performance decreases significantly under the AUC metric after removing unreached mutants. Therefore, they recommend that researchers remove uncovered  mutants when reporting predictive performance. Moreover, Guerrero-Contreras et al. \cite{guerrero2025explainable} and Balderas-Díaz et al. \cite{balderas2025analysis} specifically examined the interpretability of predictive models and the issue of class imbalance in predictive mutation testing, respectively.

\subsection{Predictive Mutation Testing at Different Granularities}

Let $T$ denote the test suite $T$=\{$t_1$, $t_2$, $\cdots$, $t_k$\}, and let $M$ denote the mutant set $M$=\{$m_1$, $m_2$, $\cdots$, $m_n$\}. Let $MS$ denote the actual mutation score of the test suite $T$ and let $MS’$ represent the predicted mutation score of $T$. Let $Mat$ denote the predicted kill matrix. 

In predictive mutation testing, previous work \cite{zhang2018predictive, mao2019extensive, zhang2020cbua, aghamohammadi2021ensemble, guerrero2025explainable, balderas2025analysis} predicts whether each mutant in the mutant set $M$ will be killed or not. These studies are at a \textit{coarse-grained} level. Based on the prediction results for each mutant in $M$, the predicted mutation score $MS’$ of $T$ can be calculated.

\begin{figure}[H]
\centering

\begin{minipage}[b]{0.48\textwidth}
\centering
\begin{lstlisting}[style=javastyle]
final static byte BYTE_LF = (byte) '\n';
private final static int[] _icUTF8 = CharTypes.getInputCodeUtf8();
protected final static int[] _icLatin1 = CharTypes.getInputCodeLatin1();
protected ObjectCodec _objectCodec;
final protected ByteQuadsCanonicalizer _symbols;
protected int[] _quadBuffer = new int[16];
protected boolean _tokenIncomplete;
private int _quad1;
\end{lstlisting}
{\small (a) Field declarations}
\end{minipage}%
\hfill
\begin{minipage}[b]{0.48\textwidth}
\centering
\begin{lstlisting}[style=javastyle]
static {
    StringBuilder sb = new StringBuilder(STD_BASE64_ALPHABET);
    sb.setCharAt(sb.indexOf("+"), '-');
    sb.setCharAt(sb.indexOf("/"), '_');
    MODIFIED_FOR_URL = new Base64Variant("MODIFIED-FOR-URL", sb.toString(), false, Base64Variant.PADDING_CHAR_NONE, Integer.MAX_VALUE);
}
\end{lstlisting}
{\small (b) Static initializer blocks}
\end{minipage}

\vspace{1.5em}

\begin{minipage}[b]{0.48\textwidth}
\centering
\begin{lstlisting}[style=javastyle]
{
    setDefaultFullDetail(true);
    setArrayContentDetail(true);
    setUseClassName(true);
    setUseShortClassName(true);
    setUseIdentityHashCode(false);
    setContentStart("(");
    setContentEnd(")");
    setFieldSeparator(", ");
    setArrayStart("[");
    setArrayEnd("]");
}
\end{lstlisting}
{\small (c) Instance initializer blocks}
\end{minipage}%
\hfill
\begin{minipage}[b]{0.48\textwidth}
\centering
\begin{lstlisting}[style=javastyle]
public enum NumberType {
    INT, LONG, BIG_INTEGER, FLOAT, DOUBLE, BIG_DECIMAL
};
public class FastDatePrinter implements DatePrinter, Serializable {
    private interface NumberRule extends Rule {
        void appendTo(StringBuffer buffer, int value);
    }
}
\end{lstlisting}
{\small (d) Definitions of inner classes, enums, or interfaces}
\end{minipage}

\caption{Code snippet of outside-method statements.}
\label{fig:java_declarations}
\end{figure}

In contrast, recent studies \cite{kim2022predictive, jain2023contextual, zhao2024spotting} predict the kill matrix, $Mat$. Specifically, for each mutant-test pair ($m_i$, $t_i$), these studies predict whether the test case $t_i$ will kill the mutant $m_i$. These approaches are at a \textit{fine-grained} level. Based on the predicted kill matrix $Mat$, the $MS’$ of $T$ can be calculated. In addition, whether each mutant is killed by $T$ can also be obtained from $Mat$. 

\subsection{Outside-Method Mutation-Test Pairs}

Three recent studies focus on predicting the kill matrix. Kim et al. \cite{kim2022predictive} introduced Seshat, which leverages the association between the names of source and test methods. Seshat uses six features, including the names and parameter types of both methods, and employs a deep learning model with a comparison layer to capture the semantic similarity between these names. However, Seshat cannot predict for mutations occurring outside source methods, as it relies on method names. 

Later, two other approaches, MutationBERT  \cite{jain2023contextual} and SODA \cite{zhao2024spotting}, were proposed to predict the kill matrix. MutationBERT and SODA collect features such as method signatures and contextual information, i.e., statements surrounding a mutated statement within the same source method. Because these features rely heavily on source method information, MutationBERT and SODA still cannot make predictions for mutations occurring outside source methods. Outside-method statements typically include field declarations, static initializer blocks, instance initializer blocks, and definitions of inner classes, enums, or interfaces. These outside-method statements play an important role in defining class structure and behavior \cite{alves1999formal}. Fig. \ref{fig:java_declarations} presents code snippets of outside-method statements from the projects used in this study.

\begin{table}[htbp]
\centering
\scriptsize
\caption{Number of Mutant-Test Pairs from the Projects Used in This Study}
\label{tab:mutant-test_pairs}
\begin{tabular}{ccccc}
\toprule
\multirow{3.5}{*}{\textbf{Project Versions}} & \multicolumn{4}{c}{\textbf{Mutant-Test Pairs}} \\
\cmidrule(lr){2-5}
 & \multirow{2}{*}{\textbf{All}} & \textbf{Inside-} & \textbf{Outside-} & \textbf{Outside-Method} \\
 & ~ & \textbf{Method} & \textbf{Method} & \textbf{/ All \%} \\
\midrule
Chart\_1 & 1,104,410 & 966,273 & 138,137 & 12.51 \\
Chart\_5 & 968,065 & 839,446 & 128,619 & 13.29 \\
Chart\_10 & 855,517 & 736,020 & 119,497 & 13.97 \\
Chart\_15 & 845,662 & 727,878 & 117,784 & 13.93 \\
Chart\_20 & 770,726 & 660,518 & 110,208 & 14.30 \\
Chart\_25 & 765,487 & 655,364 & 110,123 & 14.39 \\
\midrule
JacksonCore\_25 & 681,309 & 528,887 & 152,422 & 22.37 \\
JacksonCore\_20 & 495,715 & 392,026 & 103,689 & 20.92 \\
JacksonCore\_15 & 407,294 & 314,306 & 92,988 & 22.83 \\
JacksonCore\_10 & 400,214 & 308,935 & 91,279 & 22.81 \\
JacksonCore\_5 & 272,563 & 206,189 & 66,374 & \textbf{24.35} \\
JacksonCore\_1 & 198,366 & 150,618 & 47,748 & 24.07 \\
\midrule
Gson\_15 & 413,300 & 380,421 & 32,879 & 7.96 \\
Gson\_10 & 397,077 & 365,494 & 31,583 & 7.95 \\
Gson\_5 & 389,760 & 358,491 & 31,269 & 8.02 \\
Gson\_1 & 226,176 & 216,460 & 9,716 & 4.30 \\
\midrule
Lang\_1 & 223,948 & 183,722 & 40,226 & 17.96 \\
Lang\_10 & 215,930 & 176,323 & 39,607 & 18.34 \\
Lang\_20 & 178,173 & 144,052 & 34,121 & 19.15 \\
Lang\_30 & 171,672 & 143,659 & 28,013 & 16.32 \\
Lang\_40 & 170,907 & 143,240 & 27,667 & 16.19 \\
Lang\_50 & 175,801 & 150,559 & 25,242 & 14.36 \\
Lang\_60 & 147,244 & 125,697 & 21,547 & 14.63 \\
\midrule
Cli\_30 & 58,966 & 58,966 & 0 & 0.0 \\
Cli\_20 & 26,540 & 26,540 & 0 & 0.0 \\
Cli\_10 & 22,062 & 22,062 & 0 & 0.0 \\
Cli\_1 & 18,017 & 18,017 & 0 & 0.0 \\
\midrule
Csv\_15 & 48,824 & 45,343 & 3,481 & 7.13 \\
Csv\_10 & 26,393 & 24,923 & 1,470 & 5.57 \\
Csv\_5 & 24,342 & 22,968 & 1,374 & 5.64 \\
Csv\_1 & 8,951 & 8,575 & 376 & 4.20 \\
\bottomrule
\end{tabular}
\end{table}

In real-world projects, mutations occurring outside source methods can represent a significant portion. Table \ref{tab:mutant-test_pairs} provides a breakdown of mutant-test pairs from the projects used in our study, distinguishing between inside-method and outside-method mutants. The data in Table \ref{tab:mutant-test_pairs} are collected from mutants generated by Major \cite{just2014major}, a mutation testing tool commonly used in fine-grained predictive mutation testing approaches \cite{kim2022predictive, jain2023contextual, zhao2024spotting}. For most subjects, the proportion of outside-method mutants exceeds 10\%. Notably, in the JacksonCore\_5 project, outside-method mutants account for 24.35\% of the total mutants (displayed in bold in Table \ref{tab:mutant-test_pairs}). This highlights the practical need for a fine-grained predictive mutation testing approach that can handle both outside-method and inside-method mutants. Without this capability, the accuracy of predictive mutation testing could be significantly impacted. This challenge motivated the development of our WITNESS.

\section{Approach}

In this section, we provide the detailed steps of WITNESS for predicting the kill matrix. We also present approaches to conduct test case prioritization using the predicted kill matrix.

\subsection{Overview}

\begin{figure}[htbp]
    \centering
        \includegraphics[scale=0.16]{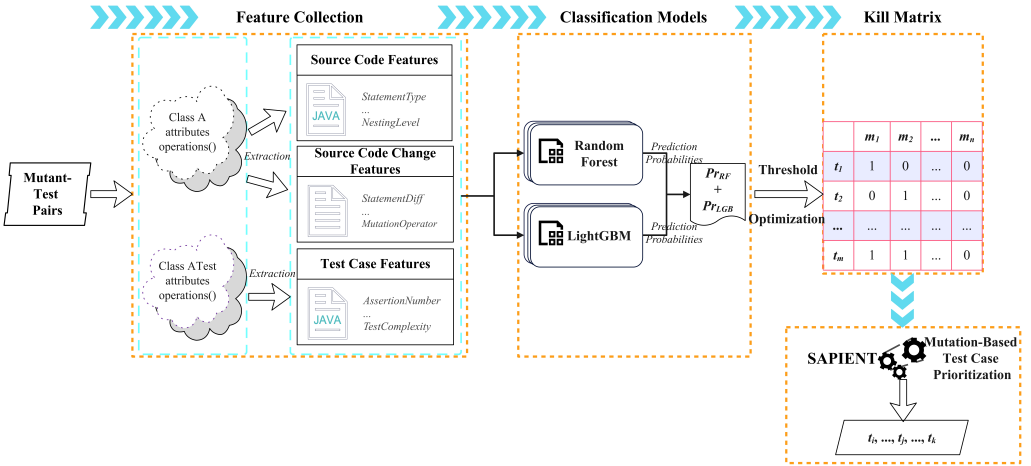}
    \caption{Workflow of WITNESS and its downstream task, SAPIENT.}
    \label{fig:WITNESS}
\end{figure}

Fig. \ref{fig:WITNESS} presents the workflow of WITNESS and the downstream task conducted in this paper, specifically, mutation-based test case prioritization. For each mutant-test pair, we collect features associated with the mutant and the test case. We then use machine learning models to predict whether the test case could kill the mutant. After combining all the prediction results of the mutant-test pairs, the full kill matrix can be obtained. Using this predicted kill matrix, the killing results of each mutant can be obtained, allowing for the calculation of a predicted mutation score. This score can be used to evaluate test suite effectiveness.

The predicted kill matrix provides an opportunity to directly perform downstream tasks. For our application study, we focus on test case prioritization (TCP), which aims to execute test cases that detect defects earlier, allowing for quicker defect detection. Building on this, after predicting the kill matrix using WITNESS, we perform mutation-based test case prioritization, which we name SAPIENT (te\underline{S}t c\underline{A}se \underline{P}rioritizat\underline{I}on bas\underline{E}d o\underline{N} WI\underline{T}NESS).

\subsection{Feature Collection}

WITNESS gathers a total of 21 features, which are categorized into three main groups: source code, source code changes, and test cases. Since predicting the kill matrix involves pairs of mutants and test cases, the collected features include information about the source code, the source code changes, and the test case. We believe that features from these three categories are sufficient to perform the prediction task. We provide detailed explanations of each feature selected within these categories.

\subsubsection{Source Code}

\begin{itemize}
\item \textit{StatementType:} This refers to the type of a mutated statement. For example, mutations may occur in an \textit{assignment} statement, an \textit{‘if’} or \textit{‘while’} condition, or a \textit{return} statement.

\item \textit{ParentContextType:} This refers to the type of the parent context. For example, if the mutated statement is an assignment statement and lies within an \textit{if} block, then the parent context type is \textit{if}.

\item \textit{LinesInMethod:} This refers to the number of lines of code in a source method. If the mutation occurs outside source methods, the value of this feature is 0.

\item \textit{SourceComplexity:} This refers to the McCabe complexity of a source method. If the mutation occurs outside source methods, the value of this feature is 0.

\item \textit{Call:} This indicates the number of source methods that are called by a source method. If the mutation occurs outside source methods, the value of this feature is 0. A method that calls many others may have a broader behavioral impact, thus a mutation in it can potentially affect downstream behavior and be more detectable by test cases.

\item \textit{Callby:} This indicates the number of source methods that call a source method. If the mutation occurs outside source methods, the value of this feature is 0. If a source method is widely invoked by other methods, mutations in it are more likely to be executed in various contexts, making them easier to detect during testing.

\item \textit{ConditionalBlockLOC:} This refers to the number of lines of code in the current conditional block. For example, if an \textit{while} block contains three lines of code, the value of this feature is 3.

\item \textit{ConditionalBlockCount:} This refers to the number of conditional blocks within the current conditional block. For example, if a \textit{for} block contains two \textit{while} blocks, then the value of this feature is 2.

\item \textit{NestingLevel:} This indicates the nesting level of mutated statements within a conditional block. For example, if a mutated statement is inside a \textit{while} block that is nested within an if block, the nesting level of the mutated statement is 2.

\item \textit{OccurringCount:} This refers to the total number of times variables in a conditional statement appear in the conditional block. For example, if a \textit{while} condition is mutated, the variables in the condition are first extracted, and their occurrences are then summed over the corresponding \textit{while} code block. The intuition is that if the occurrences are higher, the variables in the conditional statement are used more frequently, making mutations in the conditional statement easier to detect.

\item \textit{HasReturnOrThrow:} This indicates whether the current conditional block contains \textit{return} or \textit{throw} statements. If not, but a variable is assigned within the current conditional block and that variable is returned in the same class, then the conditional block is still considered to have a \textit{return} statement. The intuition is that if a conditional block returns a value or throws an exception, its behavior is more likely to be observed or trigger a failure during test case execution.

\item \textit{DeclaredVariableType:} This refers to the type of a variable being declared.

\item \textit{VariableIsFinalNew:} This indicates whether the declared variable is marked as \textit{final} or is assigned a newly created object using the \textit{new} keyword.

The features \textit{ConditionalBlockLOC}, \textit{ConditionalBlockCount}, and \textit{HasReturnOrThrow} apply only to statements where mutations occur on conditional statements (e.g., \textit{if} conditions).

\end{itemize}

\subsubsection{Source Code Change}

\begin{itemize}
\item \textit{MutationOperator:} This refers to the type of mutation operators used. For example, the Arithmetic Operator Replacement (AOR) mutation operator could replace $'a * 3’$ with $'a / 3'$.

\item \textit{StatementDiff:} This refers to the changes in a line of code before and after mutation. For example, if the condition changes from $'a <= b'$ to $'a >= b'$, then the value of this feature would be [$'<='$, $'>=’$].

\item \textit{SkeletonModification:} This refers to the structural changes in conditional statements. For example, if the condition $'a == b'$ is changed to \textit{false}, then the feature would be [$expr_1$ == $expr_2$, $expr$]. If the condition $'a == b'$ is changed to $'a$ != $b'$, the feature becomes [$expr_1$ == $expr_2$, $expr_1$ != $expr_2$]. If the mutation occurs outside conditional statements, the value of this feature is empty.
\end{itemize}

\subsubsection{Test Case}

\begin{itemize}
\item \textit{HitsNumber:} This refers to the number of times the mutated statement is executed by one test case. This feature is the only dynamic feature.

\item \textit{AssertionNumber:} This refers to the number of assertion statements—such as \textit{assertEquals}() and \textit{fail}()—in a test method.

\item \textit{HasThrow:} This indicates whether a test method throws exceptions. The value of this feature has only two possible values: ‘‘throws’’ if the test method throws exceptions; otherwise, an empty string. The intuition is that if a test case catches an exception, it may be handling an exception thrown by the executed code.

\item \textit{LinesInTestCase:} This refers to the number of lines of code in a test method.

\item \textit{TestComplexity:} This refers to the McCabe complexity of a test method.
\end{itemize}

\subsubsection{Discussion}

The feature-extraction process builds on and is inspired by prior findings \cite{visser2016makes, du2023kill, zhang2018predictive, mao2019extensive, aghamohammadi2021ensemble, kim2022predictive, jain2023contextual, zhao2024spotting}. Feature expansion is guided by existing features (e.g., \textit{TestComplexity} was inspired by \textit{SourceComplexity}) and by considerations related to outside-method mutant–test pairs. While the features used by WITNESS share some commonalities with coarse-grained predictive mutation testing approaches \cite{zhang2018predictive, mao2019extensive, aghamohammadi2021ensemble}, the overlapping features account for only a small proportion. For example, there are 5 features (\textit{StatementType}, \textit{LinesInMethod}, \textit{SourceComplexity}, \textit{MutationOperator}, \textit{AssertionNumber}) that WITNESS shares with PMT. In contrast, the features used by WITNESS differ significantly from those in coarse-grained predictive mutation testing approaches, particularly in the two features: \textit{StatementDiff} and \textit{SkeletonModification}. These two features include information from before and after the mutation, which can be used to distinguish whether a test case kills different mutants generated from the same source code. 

Although many prior studies \cite{madeyski2015process, moser2008comparative, menzies2010defect, shi2022software, pizzoleto2019systematic, wang2016amalgam+, rahman2018improving, youm2017improved, wang2018there, kang2022detecting, yedida2023find, ge2024machine} use static features for prediction tasks in software testing and analysis, the features collected in WITNESS differ from those in these studies. Specifically, in software defect prediction using static features \cite{madeyski2015process, moser2008comparative, menzies2010defect} —such as cyclomatic complexity or code churn—the features are extracted from source code, and test cases are neither executed nor required. For static analysis-based defect localization \cite{pizzoleto2019systematic, wang2016amalgam+, rahman2018improving, youm2017improved}, the features may be collected from code structure and bug reports, and do not use test case metrics. While features for prediction tasks in software testing and analysis often share common elements, they are adapted and specialized for each specific task, resulting in differences in the final feature set. For example, in machine learning-based actionable warning identification \cite{wang2018there, kang2022detecting, yedida2023find, ge2024machine}, in addition to features related to code characteristics, the features typically include warning characteristics. Although several features in WITNESS overlap with those in prior studies, it also includes features tailored to determining whether a test case kills a mutant, and its predictions are based on the interplay among three types of features: source code, source code changes, and test cases.

\subsection{Classification Models}

In predicting the kill matrix, which consists of results from mutant-test pairs, determining whether a test case kills a mutant is a binary classification task. We employ two classification-based machine-learning models, whose implementations natively support categorical features, to predict the kill matrix.

We first use the Random Forest model since it is a commonly used classification technique. PMT \cite{zhang2018predictive} uses a Random Forest to predict whether a mutant will be killed. The Random Forest model initially generates many trees from bootstrapped training samples (a random sample with replacement) and then combines predictions from all trees through majority voting for classification tasks. Mathematically, consider a Random Forest model with $B$ trees. Let each tree $b$ make a prediction $\hat{f}_b(x)$ for an input $x$. The Random Forest prediction for classification, $\hat{f}_RF(x)$, is:

\begin{equation}
\hat{f}_{RF}(x) = \text{mode}(\hat{f}_1(x), \hat{f}_2(x), \ldots, \hat{f}_B(x))
\end{equation}

The second model we use is LightGBM \cite{ke2017lightgbm}. It is a boosting method that efficiently performs classification tasks. LightGBM builds and combines a sequence of decision trees into a strong learner that optimizes an objective function. The formula to update the LightGBM model at each iteration can be expressed as follows:

\begin{equation}
F_m(x) = F_{m-1}(x) + r_m h_m(x)
\end{equation}
where $F_m (x)$ is the model at iteration $m$, $F_{(m-1)} (x)$ is the model from the previous iteration, $h_m (x)$ is the new tree added at iteration $m$, and $r_m$ is the learning rate applied to the new tree, which controls the impact of each individual tree and prevents overfitting. 

We train Random Forest and LightGBM models separately, then perform evaluations on the test set. We combine the predicted probabilities of the two models by adopting the average value of the predicted probabilities. Combining prediction probabilities from different models helps mitigate the limitations of any single model and provides better predictive performance. This process can be denoted by the following formula:

\begin{equation}
P_{combined} = \frac{P_{RF} + P_{LGBM}}{2}
\end{equation}
where  $P_{RF}$  and  $P_{LGBM}$  are the predicted probabilities from the Random Forest and LightGBM models, respectively.

Since some machine learning models are sensitive to the scale of the data \cite{ozdemir2018feature}, we perform feature scaling, also known as data normalization. Data normalization allows machine learning models to learn optimally and prevents bias towards larger-scaled features \cite{ozdemir2018feature}. Our chosen method for normalization is z-score standardization, recalibrating features to have a mean of 0 and a standard deviation of 1. We apply the formula $z = \frac{(x - \mu)}{\sigma}$, where $x$ is a feature value, $\mu$ is the mean of the features, and $\sigma$ is their standard deviation. We perform this data normalization for the numerical features in the feature set of WITNESS.

For mutant-test pairs, those in which a test case kills a mutant typically represent the minority class \cite{jain2023contextual}. Consequently, the optimal threshold may not be 0.5. Therefore, we further determine the optimal threshold. Note that, MutationBERT varies the thresholds from 0.01 to 1.0 in increments of 0.01 to obtain the optimal threshold of 0.25 for predicting whether a mutant will be killed by the whole test suite or not.

\subsection{Threshold Optimization}

Based on the prediction probabilities calculated from two machine learning models, we determine the optimal threshold to better classify the killing results of mutant-test pairs. To evaluate the effectiveness of the predictive performance in predicting the kill matrix, a confusion matrix is used, as shown in Table \ref{table:confusion_matrix}.

\begin{table}[htbp]
\centering
\scriptsize
\caption{Confusion Matrix}
\label{table:confusion_matrix}
\begin{tabular}{l c c}
\toprule
 & \textbf{Predict Killed} & \textbf{Predict Survived} \\ 
\midrule
\textbf{Actual Killed} & TP & FN \\ 
\textbf{Actual Survived} & FP & TN \\ 
\bottomrule
\end{tabular}
\end{table}

\textit{Precision} measures the accuracy of the positive predictions made by the predictive model and is calculated as follows:

\begin{equation}
Precision = \frac{TP}{TP + FP}
\end{equation}

\textit{Recall}, also known as the \textit{True Positive Rate} (TPR), measures the proportion of actual positives that are correctly identified by the model. Mathematically, it is calculated as follows:

\begin{equation}
Recall = \frac{TP}{TP + FN}
\end{equation}

The \textit{False Positive Rate} (FPR) measures the proportion of actual negatives that are incorrectly identified as positives by the predictive model. Mathematically, FPR is calculated as follows: 

\begin{equation}
FPR = \frac{FP}{FP + TN}
\end{equation}

The \textit{F1-score} is the harmonic mean of Precision and Recall, calculated by: 

\begin{equation}
F1 = \frac{2 \times Precision \times Recall}{Precision + Recall}
\end{equation}

The \textit{J statistic} \cite{youden1950index}, also known as Youden’s index, evaluates the performance of a classification model. Mathematically, it is represented as follows: 

\begin{equation}
J = TPR - FPR
\end{equation}

We use the \textit{Precision-Recall curve} \cite{harper1978evaluation} and the \textit{ROC (Receiver Operating Characteristic) curve} \cite{zweig1993receiver} to determine the optimal decision threshold for classification. The Precision-Recall curve is created by plotting Precision on the y-axis and Recall on the x-axis. To generate the curve, the threshold of the classification model is varied. The ROC curve is created by plotting the TPR on the y-axis against the FPR on the x-axis at various threshold settings.

We determine the optimal threshold from the validation set data by exploring thresholds within the interval [0.05, 0.5]. Specifically, we start at $\theta$=0.05 and increase by increments of $\Delta \theta$=0.05, ending at $\theta$=0.5. The ending at 0.5 is because the ratio of test cases that kill mutants in the mutant-test pairs is typically the minority class; having the threshold exceed 0.5 may lead to noticeably more mutant-test pairs being classified as the test case killing the mutant. This procedure results in a total of 10 potential thresholds: $\theta_1$=0.05, $\theta_2$=0.1, $\theta_3$=0.15, …, $\theta_{10}$=0.5. 

For each threshold $\theta_i$, we calculate the F1-score, $F1(\theta_i)$, from the Precision-Recall curve and the J statistic,  $J(\theta_i)$, from the ROC curve. Both of these values undergo z-score standardization to ensure comparability and account for variance among the values.

We then compute the sum of these normalized values for each threshold:

\begin{equation}
S(\theta_i) = F1'(\theta_i) + J'(\theta_i)
\end{equation}
 
Finally, we select the threshold $\theta^*$ that maximizes $S(\theta_i)$.

\begin{equation}
\theta^* = \arg \max_{\theta_i} S(\theta_i)
\end{equation}

This $\theta^*$ is considered the optimal threshold. 

Once the threshold is determined, if the predicted probability is greater than or equal to this threshold, the prediction for a mutant-test pair is set as \textit{‘the test case kills the mutant’}; otherwise, it is marked as \textit{‘the test case cannot kill the mutant’}. 

\subsection{SAPIENT}

The kill matrix is generated by aggregating the predicted results of mutant-test pairs. We perform mutation-based test case prioritization \cite{shin2019empirical} using the predicted kill matrix. Test case prioritization reorders test cases in a test suite so that the test cases that detect defects are executed earlier to improve the speed of defect detection. Given a test suite $T$, let $PT$ be the set of permutations of $T$, and let $f$ denote the function $PT$ $\rightarrow$ $\mathbb{R}$, which maps $PT$ to real numbers. The problem of test case prioritization is to find \( T' \)$\in$\( PT \) such that ($\forall$\( T'' \))(\( T'' \)$\in$\( PT \))(\( T'' \)$\neq$\( T' \))(\( f(T') \)$\ge$\( f(T'') \)) \cite{rothermel2001prioritizing}. In this definition,  $PT$ represents the set of all possible permutations of a test suite. The function $f$ maps a permutation of the test suite to an \textit{award value}, assuming that higher award values are preferable to lower ones. The objective is to find the permutation with the highest award value.

In this paper, we use the \textit{Total} and the \textit{Additional} approaches \cite{rothermel2001prioritizing} to perform test case prioritization. The Total approach sorts test cases based on the number of killed mutants (test case’s mutation score) in decreasing order. If several test cases kill the same number of mutants, we then randomly select one of the test cases.

The Additional approach iteratively selects a test case that maximizes the number of additionally killed mutants, namely, mutants that were not previously killed. If several test cases kill the same number of additional mutants, then one of the test cases is randomly selected. In the process of test case prioritization, the Additional approach may encounter situations where the remaining test cases cannot kill additional mutants. Similar to previous study \cite{rothermel2001prioritizing}, we reset the mutation scores to their initial values and reapply the Additional approach for the remaining test cases.

For the Additional approach, it is necessary to use the kill matrix to perform test case prioritization. This approach requires calculating the accumulated number of mutants that are killed by multiple test cases. Since a mutant can be killed by multiple test cases, the accumulated number of mutants killed by multiple test cases cannot simply be calculated by adding the number of mutants each test case kills. The accumulated number of mutants killed by multiple test cases can be accurately derived from the kill matrix, enabling direct mutation-based test case prioritization.

\section{Experimental Setup}

This section presents the settings of the experiments conducted to evaluate WITNESS and three baseline approaches. 

\subsection{Research Questions}

We compare the results of WITNESS with three baseline approaches that predict the kill matrix: Seshat, MutationBERT, and SODA. To evaluate these approaches, we pose the following six research questions:

~\\
\textbf{RQ1:} (\textit{Effectiveness}). \textit{How effective is WITNESS at predicting the kill matrix?}

We evaluate the effectiveness of WITNESS and the baselines across three scenarios: same-version, cross-version, and cross-project. In each scenario, we assess the results in terms of kill matrix prediction, mutant killing prediction, and mutation score prediction. Mutant killing prediction (the killing results of each mutant by the entire test suite) and mutation score prediction are based on the predicted kill matrix, offering different perspectives on the effectiveness of the predicted kill matrix generated by various predictive approaches.

We compare the effectiveness of WITNESS with the baselines in predicting mutant-test pairs where mutations occur within source methods. Additionally, we present the results of WITNESS in predicting all mutant-test pairs, including those where mutations occur outside source methods.

~\\
\textbf{RQ2:} (\textit{Predictive Performance for Different Mutant Killing Reasons}). \textit{How well does WITNESS perform when categorized by different mutant killing reasons?}

Du et al. \cite{du2023kill} classify test case failures into three categories: test oracle failure, source-code oracle failure, and exogenous crashes. Similarly, Major \cite{just2014major} classifies test case failures into assertion failure, exception, and timeout. Therefore, mutant killing can occur for different reasons. The original purpose of mutation testing is to evaluate the effectiveness of test suites. Despite the various reasons for mutant killing, the effectiveness of test suites is directly linked to assertion failures (test oracle failures). Specifically, mutants are directly killed by test cases when assertion failures occur.

Using the models trained by WITNESS and the baselines, we compare the predictive performance for these three distinct types of killing reasons: assertion failure, exception, and timeout, with a particular focus on assertion failures. The evaluation of predictive performance for different mutant killing reasons is based on the predicted kill matrices generated by different predictive approaches.

~\\
\textbf{RQ3:} (\textit{Prediction of Outcomes for Fewer-Covered Mutants}). \textit{How effective is WITNESS in predicting the results of fewer-covered mutants?}

Fewer-covered mutants are those covered by fewer test cases among all test cases that could cover mutants. For example, a mutant might be covered by just one or two test cases. As fewer test cases can cover them, the number of formed mutant-test pairs for these mutants is smaller compared to mutants that are covered by many test cases.

Fewer-covered mutants indicate areas of the program under test that are less likely to be exercised by existing test cases. Identifying such mutants can reveal test suite inadequacies and guide the generation of targeted test cases to improve test suite effectiveness. Additionally, since the predicted results of mutants are aggregated from the predicted results of mutant-test pairs, incorrect predictions for mutant-test pairs associated with fewer-covered mutants can easily lead to erroneous predictions for these mutants. We compare the predictive performance of WITNESS in predicting the results of fewer-covered mutants against that of the baselines.

~\\
\textbf{RQ4:} (\textit{Contribution of Features}). \textit{Which feature contributes the most to the predictive performance in fine-grained predictive mutation testing?}

Previous coarse-grained predictive mutation testing studies \cite{zhang2018predictive, mao2019extensive, aghamohammadi2021ensemble} analyzed feature importance. However, the three studies \cite{kim2022predictive, jain2023contextual, zhao2024spotting} that use deep learning to predict the kill matrix do not present feature importance. 

The adoption of machine learning models for predicting the kill matrix, rather than using deep learning models, allows for an analysis of feature importance. We use feature importance to interpret the decision-making process involved in determining whether a test case will kill a mutant. The importance of different features helps guide further research in fine-grained predictive mutation testing.

~\\
\textbf{RQ5:} (\textit{Efficiency}). \textit{How efficient is WITNESS compared to the baseline approaches?}

Predictive mutation testing is proposed to reduce the high costs of actual mutation testing. Thus, it is important to evaluate the efficiency of the proposed approaches. We measure the inference time for Seshat, MutationBERT, SODA, and WITNESS, respectively, and then calculate the speedup of WITNESS relative to the baselines.

~\\
\textbf{RQ6:} (\textit{Prioritization of Test Cases}). \textit{How well is the predicted kill matrix by WITNESS used for test case prioritization?}

Mutation-based test case prioritization can be conducted using the kill matrix. Using the actual kill matrix, we obtain the true test case prioritization results. These prioritization results can also be derived from the predicted kill matrix generated by WITNESS and the three baselines. We then compare the predicted test case prioritization results from the four predictive approaches to see which are closest to the actual results. The closer the results, the more similar the predicted kill matrix is to the actual kill matrix.

\subsection{Baselines}

Seshat is the first fine-grained predictive mutation testing approach. Specifically, Seshat’s evaluation compared its performance with PMT and demonstrated higher predictive performance. Therefore, we do not include PMT as a baseline to avoid redundancy. Nevertheless, the supplementary material \cite{lu2024supplementary} provides simple comparisons with PMT, showing that WITNESS achieves better predictive performance.

MutationBERT claims to outperform Seshat in predictive performance but has lower efficiency than Seshat. However, the evaluation of MutationBERT is insufficient, as it includes only two experiments, making it unclear how its predictive performance would hold across a more comprehensive set of experiments. Compared to Seshat and MutationBERT, SODA demonstrates superior predictive performance in its evaluation. However, the evaluation of SODA is also based primarily on just two experiments, which remains insufficient in scope. The predictive performance of MutationBERT and SODA under varied dataset splits and a larger number of experiments remains unclear.

Overall, we select fine-grained predictive mutation testing approaches, namely Seshat, MutationBERT, and SODA, as our baselines. We use the code provided by the baseline methods and re-train the models for different predictive approaches. We use the code provided by the baseline methods to train predictive models and evaluate their performance.

\subsection{Dataset}

\subsubsection{Projects.} We conduct our experiments using Defects4J 2.0.0 \cite{just2014defects4j}, selecting six projects commonly used by baselines. Table \ref{tab:project_info} provides detailed information on the project versions used in the six Defects4J projects. The transition from dark purple to light purple indicates the evolution of project versions throughout time. Here, the versions refer to identifiers of selected bugs in Defects4J, rather than to project versions in the project’s progression. \textit{For the four projects—Csv, Cli, Gson, and JacksonCore—the version numbers increase over time, whereas for the two projects—Lang and Chart—the version numbers decrease across time.} 

\begin{table}[tbp]
\centering
\scriptsize
\caption{Information of Projects in Defects4J}
\label{tab:project_info}
\begin{threeparttable}
\begin{tabular}{cccccc}
\toprule
\textbf{Project ID} & \textbf{Version} & \textbf{LoC} & \textbf{\# Test Case} & \textbf{Date} \\
\midrule
\multirow{6}{*}{Chart} & \cellcolor{Purple7} 1 & \cellcolor{Purple7} 96,382 & \cellcolor{Purple7} 2,193 & \cellcolor{Purple7} 2010-02-09 \\
 & \cellcolor{Purple6} 5 & \cellcolor{Purple6} 89,347 & \cellcolor{Purple6} 2,033 & \cellcolor{Purple6} 2008-11-24 \\
 & \cellcolor{Purple5} 10 & \cellcolor{Purple5} 84,482 & \cellcolor{Purple5} 1,805 & \cellcolor{Purple5} 2008-06-10 \\
 & \cellcolor{Purple4} 15 & \cellcolor{Purple4} 84,134 & \cellcolor{Purple4} 1,782 & \cellcolor{Purple4} 2008-03-19 \\
 & \cellcolor{Purple3} 20 & \cellcolor{Purple3} 80,508 & \cellcolor{Purple3} 1,651 & \cellcolor{Purple3} 2007-10-08 \\
 & \cellcolor{Purple2} 25 & \cellcolor{Purple2} 79,823 & \cellcolor{Purple2} 1,617 & \cellcolor{Purple2} 2007-08-28 \\
\midrule
\multirow{6}{*}{JacksonCore} & \cellcolor{Purple7} 25 & \cellcolor{Purple7} 25,218 & \cellcolor{Purple7} 573 & \cellcolor{Purple7} 2019-01-16 \\
 & \cellcolor{Purple6} 20 & \cellcolor{Purple6} 21,480 & \cellcolor{Purple6} 384 & \cellcolor{Purple6} 2016-09-01 \\
 & \cellcolor{Purple5} 15 & \cellcolor{Purple5} 18,652 & \cellcolor{Purple5} 346 & \cellcolor{Purple5} 2016-03-21 \\
 & \cellcolor{Purple4} 10 & \cellcolor{Purple4} 18,930 & \cellcolor{Purple4} 330 & \cellcolor{Purple4} 2015-07-31 \\
 & \cellcolor{Purple3} 5 & \cellcolor{Purple3} 15,687 & \cellcolor{Purple3} 240 & \cellcolor{Purple3} 2014-12-07 \\
 & \cellcolor{Purple2} 1 & \cellcolor{Purple2} 15,882 & \cellcolor{Purple2} 206 & \cellcolor{Purple2} 2013-08-28 \\
\midrule
 \multirow{4}{*}{Gson} & \cellcolor{Purple7} 15 & \cellcolor{Purple7} 7,826 & \cellcolor{Purple7} 1,029 & \cellcolor{Purple7} 2017-05-31 \\ 
 & \cellcolor{Purple6} 10 & \cellcolor{Purple6} 7,693 & \cellcolor{Purple6} 996 & \cellcolor{Purple6} 2016-05-17 \\
 & \cellcolor{Purple5} 5 & \cellcolor{Purple5} 7,630 & \cellcolor{Purple5} 984 & \cellcolor{Purple5} 2016-02-02 \\
 & \cellcolor{Purple4} 1 & \cellcolor{Purple4} 5,418 & \cellcolor{Purple4} 720 & \cellcolor{Purple4} 2010-11-02 \\
\midrule
 \multirow{7}{*}{Lang} & \cellcolor{Purple7} 1 & \cellcolor{Purple7} 21,788 & \cellcolor{Purple7} 2,291 & \cellcolor{Purple7} 2013-07-26 \\
 & \cellcolor{Purple6} 10 & \cellcolor{Purple6} 20,433 & \cellcolor{Purple6} 2,198 & \cellcolor{Purple6} 2012-09-27 \\
 & \cellcolor{Purple5} 20 & \cellcolor{Purple5} 18,967 & \cellcolor{Purple5} 1,876 & \cellcolor{Purple5} 2011-07-03 \\
 & \cellcolor{Purple4} 30 & \cellcolor{Purple4} 17,660 & \cellcolor{Purple4} 1,733 & \cellcolor{Purple4} 2010-03-16 \\
 & \cellcolor{Purple3} 40 & \cellcolor{Purple3} 17,435 & \cellcolor{Purple3} 1,643 & \cellcolor{Purple3} 2009-10-22 \\
 & \cellcolor{Purple2} 50 & \cellcolor{Purple2} 17,760 & \cellcolor{Purple2} 1,720 & \cellcolor{Purple2} 2007-10-31 \\
 & \cellcolor{Purple1} 60 & \cellcolor{Purple1} 16,920 & \cellcolor{Purple1} 1,590 & \cellcolor{Purple1} 2006-10-31 \\
\midrule
 \multirow{4}{*}{Cli} & \cellcolor{Purple7} 30 & \cellcolor{Purple7} 2,497 & \cellcolor{Purple7} 354 & \cellcolor{Purple7} 2010-06-17 \\
 & \cellcolor{Purple6} 20 & \cellcolor{Purple6} 1,989 & \cellcolor{Purple6} 148 & \cellcolor{Purple6} 2008-07-28 \\
 & \cellcolor{Purple5} 10 & \cellcolor{Purple5} 2,002 & \cellcolor{Purple5} 112 & \cellcolor{Purple5} 2008-05-29 \\
 & \cellcolor{Purple4} 1 & \cellcolor{Purple4} 1,937 & \cellcolor{Purple4} 94 &  \cellcolor{Purple4} 2007-05-15 \\
\midrule
  \multirow{4}{*}{Csv} & \cellcolor{Purple7} 15 & \cellcolor{Purple7} 1,619 & \cellcolor{Purple7} 290 & \cellcolor{Purple7} 2017-12-11 \\
 & \cellcolor{Purple6} 10 & \cellcolor{Purple6} 1,276 & \cellcolor{Purple6} 200 & \cellcolor{Purple6} 2014-06-09 \\
 & \cellcolor{Purple5} 5 & \cellcolor{Purple5} 1,236 & \cellcolor{Purple5} 189 & \cellcolor{Purple5} 2014-03-13 \\
 & \cellcolor{Purple4} 1 & \cellcolor{Purple4} 806 & \cellcolor{Purple4} 54 & \cellcolor{Purple4} 2012-03-27 \\
\bottomrule
\end{tabular}
    \begin{tablenotes}
	\item The dark purple color represents the earlier versions of a project, while the light purple color represents the latest versions. 
    \end{tablenotes}
\end{threeparttable}
\end{table}

Unlike MutationBERT, which restricts itself to a single version per project, we include multiple versions from each project. The versions used in each project in our study are aligned with those in both Seshat and SODA. Using the same project versions as the baselines makes it easy to replicate their experiments using the provided preprocessed data and code. It also allows us to compare our rerun results with those reported in the baseline papers. Choosing project versions that are at least five versions apart helps ensure meaningful evaluations under the cross-version scenario. Choosing project versions that are at least five versions apart helps ensure meaningful evaluations under the cross-version scenario. This is supported by the differences in LoC, number of test cases, and version dates for each project shown in Table \ref{tab:project_info}. While we align with the project versions used in the baselines, our experimental setup incorporates distinct differences. We conduct a thorough evaluation involving significantly more experiments across different scenarios.

\subsubsection{Uncovered Mutants.} As mentioned in Section 2, Aghamohammadi et al. \cite{aghamohammadi2021ensemble} found that uncovered mutants lead to an overestimation of the results in predictive mutation testing. If a mutant is not executed by any test case within a test suite, it is considered an uncovered mutant. These uncovered mutants are guaranteed to survive, making it unnecessary to predict their results. In line with the baselines \cite{jain2023contextual, zhao2024spotting}, we predict the kill matrix and calculate the mutation score only for covered mutants.

\subsubsection{Dataset Splitting.} The baselines all divide the dataset into a training set, a validation set, and a test set. For the baselines, the validation set is used for monitoring and evaluation during training. Additionally, for MutationBERT and SODA, the validation set is also used to determine the threshold for classifying whether a mutant will be killed by a test suite. Similar to the baselines, we divide the dataset into a training set, a validation set, and a test set. The validation set in WITNESS is used for model evaluation during training and for determining the optimal threshold for classifying the killing results of mutant-test pairs.

We adopt the same approach as the baselines for preprocessing the dataset. To form the dataset comprised of mutant-test pair features, we first divide the mutants into a training set, a validation set, and a test set. We then form datasets from the features of mutant-test pairs, which are created by associating each mutant with the corresponding test cases that can reach the mutated statement. As a result, the mutants in the training set, validation set, and test set do not intersect, which avoids potential leakage between splits.

\subsubsection{Evaluation Scenarios.} Kim et al. \cite{kim2022predictive} evaluated the predictive performance of Seshat under cross-version scenario within a single project. Jain et al. \cite{jain2023contextual} evaluated MutationBERT in two experiments, corresponding to two scenarios: same-project and cross-project. Zhao et al. \cite{zhao2024spotting} evaluated SODA through mainly two experiments, covering the cross-version and cross-project scenarios. In the same-project scenario, MutationBERT divides mixed data from six bugs, with each bug coming from a different project, into a training set, a validation set, and a test set. Despite all sets containing mixed data from the six projects, MutationBERT categorizes this division as a same-project setting. We do not consider mixing data from different projects as the same-project scenario. To conduct a thorough evaluation, we evaluate WITNESS and the baselines under same-version, cross-version, and cross-project scenarios. 

The sources of the datasets and the ways of splitting them under the different scenarios are detailed below.

(1) \textit{Same-Version Scenario}


For each project version within a project, we split the training, validation, and test sets in an 80-10-10 proportion. Unlike the evaluation approach of MutationBERT in the same-project scenario, which combines data from six different projects for the training, validation, and test datasets, we maintain project-specific datasets.

(2) \textit{Cross-Version Scenario}


The prediction involves two versions in a project, and the number of experiments can be calculated using the binomial coefficient $\binom{n}{2}$, where $n$ is the total number of versions in the project. Similar to Seshat, we split the data from the earlier version into a training set and validation set with 90\% and 10\% of the whole dataset, respectively. The test set consists of the data from the later version. Later versions include mutant-test pairs not contained in earlier versions due to changes in the software. Seshat evaluates their approach exclusively under the cross-version scenario.

(3) \textit{Cross-Project Scenario}


Similar to MutationBERT and SODA, we use the latest version of a project in the cross-project scenario, as it is more representative of the project overall.

\textit{One-to-one prediction.} We use data from one project for both the training and validation sets, and data from another project as the test set. The data from the project used for training and validation is split in the same way as in the cross-version scenario. The experiments are conducted using any two of the projects. We exclude the Chart project because it has the largest number of mutant-test pairs, differing significantly from Csv (966,273 vs. 45,343). This makes prediction from Csv to Chart unsuitable for typical supervised learning tasks and leads to poor performance across predictive approaches. For the remaining five projects, the number of experiments is calculated by $P(n, r) = \frac{n!}{(n - r)!}$, where $n$ represents the total of 5 projects, and $r$ represents the 2 projects chosen.

\textit{Many-to-one prediction.} For all projects except Chart, the training and validation sets consist of mixed data from four projects, while the test set contains data from the remaining project. The training and validation sets are split in a 90\%-10\% ratio. We exclude Chart, as combining it with any three or more other projects results in a significantly larger training set, which requires substantial time and resources for training.

Since we have excluded Chart due to the large training sets that result from combining it with multiple projects, we include three additional experiments where the training and validation sets consist of data from the projects Chart, Cli, and Csv. Specifically, Chart has the largest number of mutant-test pairs, while Cli and Csv are the two projects with the smallest number of mutant-test pairs. The test sets are Lang, Gson, and JacksonCore. This setup results in three experiments where the same trained model is used to predict three different projects.

\subsection{Evaluation Metrics}

In this paper, we assess whether each mutant is predicted to be killed by a test suite, based on the predicted kill matrix. This approach is a natural way to obtain the killing results for each mutant using the predicted kill matrix. However, MutationBERT and SODA specifically train new models for the killing results of each mutant, meaning they train different models for predicting the outcomes of mutant-test pairs and individual mutants, which may exaggerate the experimental results of mutant killing prediction. For the evaluation of MutationBERT and SODA, the predicted killing result of each mutant is derived from its predicted kill matrix.

\subsubsection{Predictive Performance}

In predicting the \textit{kill matrix}, the mutant-test pairs where the test case kills the mutant are typically the minority class. Therefore, we use Precision, Recall, and F1-score to evaluate the predictive performance of the predicted kill matrix. All three metrics focus on the killed mutant-test pairs.

After obtaining the predicted kill matrix, we evaluate the predictive performance in determining \textit{whether a mutant is killed or survived}. Since survived mutants highlight the weaknesses of a test suite and can guide efforts to improve its effectiveness, we designate survived mutants as the positive class when evaluating the predictive results of each mutant under the entire test suite.

Based on the predicted kill matrix, we evaluate the predicted \textit{mutation score} achieved by a test suite using APE (Absolute Prediction Error) \cite{zhang2020cbua}. APE = $|pms - ams|$, which represents the absolute difference between the predicted mutation score ($pms$) and the actual mutation score ($ams$). The smaller the value of APE, the better the predicted mutation score. 

\subsubsection{Predictive Performance for Different Mutant Killing Reasons}

Mutant-test pairs where a test case actually kills a mutant can be categorized by mutant killing reasons. To evaluate predictive performance for different mutant killing reasons, the predictions are conducted on a sub-test set labeled to indicate that a test case kills a mutant—essentially, the actual killed mutant-test pairs in the test set. Therefore, the prediction does not include actual survived mutant-test pairs (i.e., pairs where a test case does not kill a mutant). 

Consequently, there are two outcomes in the prediction: ‘predict killed’ and ‘predict survived’ among the actual killed mutant-test pairs. In the confusion matrix, both False Positives (FP) and True Negatives (TN) are zero, resulting in a Precision of 1.0. Since the F1-score is the harmonic mean of Precision and Recall, we use Recall as the evaluation metric.

\subsubsection{Fewer-Covered Mutants}

We evaluate the predictive performance for mutants covered by fewer test cases among all test cases that could cover mutants. We adopt the F1-score to evaluate such mutants.

\subsubsection{Efficiency}

The baselines commonly report the time taken for predictions on the test set. In this paper, we focus on measuring and comparing the prediction times to evaluate the efficiency of the four approaches. 

\subsubsection{Test Case Prioritization}

To assess the effectiveness of test case prioritization results, we use the most widely adopted metric, APFD (Average Percentage of Faults Detected) \cite{khatibsyarbini2018test}:

\begin{equation}
APFD = 1 - \frac{TF_1 + TF_2 + \ldots + TF_r}{nr} + \frac{1}{2n}
\end{equation}

where $r$ represents the number of defects, $n$ denotes the number of test cases in the test suite $T$, and $TF_i$ is the ranking of the test case in the prioritized test suite that first detects the $i_{th}$ defect. 

\subsection{Implementation}

The experiments for WITNESS and the three baselines were conducted on a server equipped with an Intel(R) Xeon(R) Platinum 8268 CPU at 2.90GHz, two Nvidia Tesla V100 GPUs, 118GB of memory, and running the Ubuntu 18.04 LTS operating system. Specifically, GPUs are used to replicate the baseline experiments, whereas WITNESS runs without GPU support. The data analysis was conducted on a computer with 16GB of memory and running the macOS 15 operating system.

We use Major \cite{just2014major} for mutation testing, as it is the default mutation testing tool in Defects4J. Additionally, it is utilized in both baseline approaches. Specifically, MutationBERT and SODA exclusively uses Major in its experiments. Major includes eight mutation operators, namely: AOR (Arithmetic Operator Replacement), LOR (Logical Operator Replacement), COR (Conditional Operator Replacement), ROR (Relational Operator Replacement), SOR (Shift Operator Replacement), ORU (Operator Replacement Unary), EVR (Expression Value Replacement), LVR (Literal Value Replacement), and STD (Statement Deletion).

To collect the 21 features used in WITNESS, we utilized the ANTLR (ANother Tool for Language Recognition) \cite{parr2013definitive}, Understand \cite{understand}, and Cobertura \cite{cobertura} tools. ANTLR and Understand were used to collect static features, while Cobertura was used to collect coverage information. Given that the feature set contains categorical features, we utilize implementations of models that directly support handling categorical data. Since the Random Forest model implemented in scikit-learn \cite{pedregosa2011scikit} does not natively handle categorical data, we use the H2O \cite{h2o} Random Forest model instead. We use the default encoding scheme for handling categorical features in H2O Random Forest, which adopts \textit{enum encoding}. For the implementation of the LightGBM model, we use the lightgbm package \cite{lightgbm} in Python. Categorical features are \textit{integer-encoded} in the LightGBM package, which uses Fisher’s method \cite{fisher1958grouping} to find the optimal split for categorical features. We also leverage the scikit-learn library \cite{pedregosa2011scikit} to help process the dataset and evaluate predictive performance.

\section{Experimental Results}

In this section, we present the experimental results for the predictive performance of WITNESS and comparison with the baselines, along with an analysis of feature importance, the comparison over efficiency, and the test case prioritization results based on the predicted kill matrix.

\subsection{RQ1: Effectiveness}

We first assess the effectiveness of WITNESS and compare it with the baselines in terms of predictive performance using datasets consisting of mutant-test pairs where mutations occur within source methods. Additionally, we present the predictive results of WITNESS on all mutant-test pairs, including those where mutations occur outside source methods.

\begin{table*}[htbp]
\centering
\scriptsize
\caption{Average Predictive Performance of WITNESS and Baselines}
\begin{threeparttable}
\begin{tabular}{cccccccccccccc}
\toprule
\multirow{2}{*}{} & \multicolumn{3}{c}{\textbf{Seshat}} & \multicolumn{3}{c}{\textbf{MutationBERT}} & \multicolumn{3}{c}{\textbf{SODA}} & \multicolumn{3}{c}{\textbf{WITNESS}} & \\ 
 \cmidrule(lr){2-4}  \cmidrule(lr){5-7}  \cmidrule(lr){8-10} \cmidrule(lr){11-13}
 & \textbf{SV} & \textbf{CV} & \textbf{CP} & \textbf{SV} & \textbf{CV} & \textbf{CP} & \textbf{SV} & \textbf{CV} & \textbf{CP} & \textbf{SV} & \textbf{CV} & \textbf{CP} & \\
\midrule
                  & 0.73 & 0.75 & 0.46 & 0.53 & 0.62 & 0.38 & 0.65 & 0.70 & 0.49 & 0.67 & 0.73 & 0.46 & Precision \\
 Kill Matrix & 0.71 & 0.78 & 0.53 & 0.58 & 0.75 & 0.60 & 0.76 & 0.83 & 0.60 & 0.79 & 0.87 & 0.66 & Recall \\         
                  & 0.716 & 0.76 & 0.48 & 0.55 & 0.67 & 0.44 & 0.70 & 0.76 & 0.526 & \textbf{0.721} & \textbf{0.79} & \textbf{0.533} & F1-score \\ 
\midrule
                      & 0.58 & 0.59 & 0.34 & 0.55 & 0.66 & 0.31 & 0.70 & 0.73 & 0.43 & 0.70 & 0.74 & 0.38 & Precision \\    
Mutant Killing & 0.77 & 0.77 & 0.55 & 0.66 & 0.61 & 0.41 & 0.55 & 0.61 & 0.47 & 0.70 & 0.73 & 0.42 & Recall \\    
                      & 0.66 & 0.66 & 0.41 & 0.57 & 0.58 & 0.32 & 0.60 & 0.66 & \textbf{0.43} & \textbf{0.69} & \textbf{0.73} & 0.39 & F1-score \\ 
\midrule
Mutation Score & 11.14 & 10.01 & 16.11 & 17.75 &14.60 & 22.66 & 6.26 & 5.12 & 8.66 & \textbf{3.94} & \textbf{4.22} & \textbf{7.73} & APE \\ 
\bottomrule
\end{tabular}
    \begin{tablenotes}
	\item SV, CV, and CP represent Same-Version, Cross-Version, and Cross-Project, respectively.
    \end{tablenotes}
\end{threeparttable}
\label{tab:performance_of_witness_with_baselines}
\end{table*}

\subsubsection{Comparison with Baselines in Predicting Mutant-Test Pairs Where Mutations Occur Inside Source Methods}
~\

\textbf{\textit{Approach.}} Since the three baseline approaches are suitable only for mutants generated by mutating program elements within source methods, we limit our comparison of the three baselines to mutant-test pairs where mutations occur inside source methods. Specifically, the training of predictive models and their evaluation are based entirely on mutant-test pairs where mutations occur inside source methods. This ensures that WITNESS aligns with the baselines for the training, validation, and test sets, allowing for a fair comparison.

\begin{figure*}[htbp]
    \centering
    \subfigure[Same-Version (Kill Matrix)]{
        \includegraphics[scale=0.11]{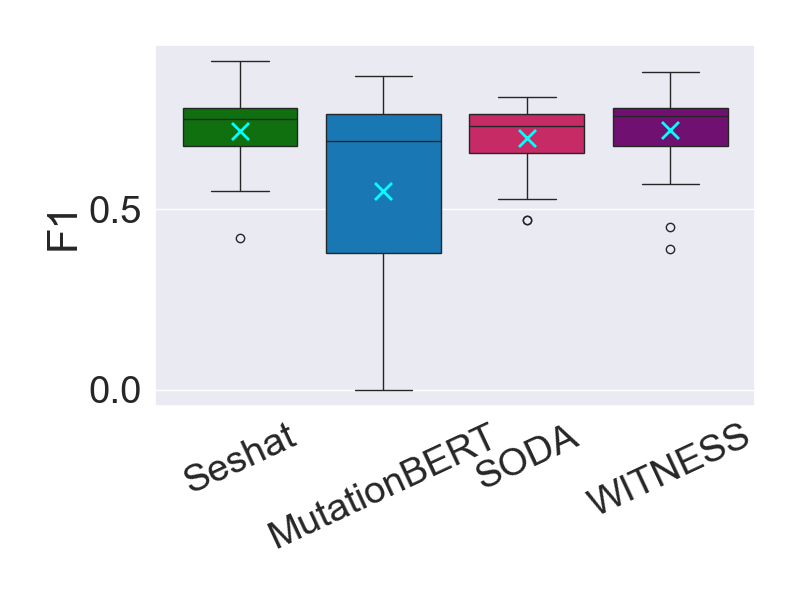}
    }
    \hfill
    \centering
    \subfigure[Cross-Version (Kill Matrix)]{
        \includegraphics[scale=0.11]{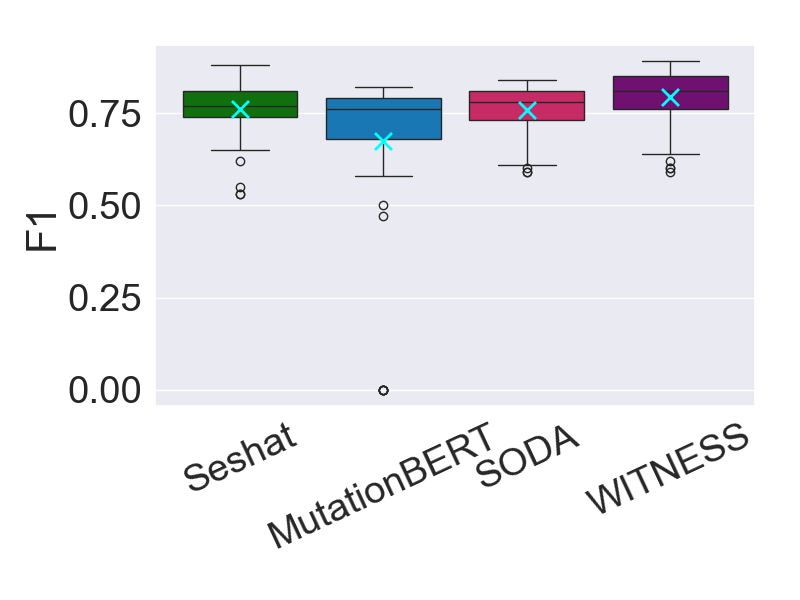}
    }
    \hfill
    \centering
    \subfigure[Cross-Project (Kill Matrix)]{
        \includegraphics[scale=0.11]{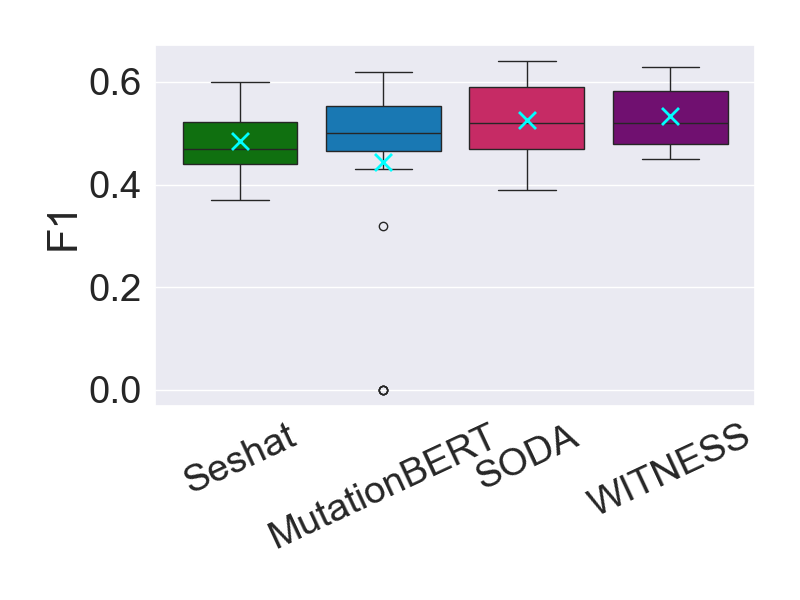}
    }
    \vfill
     \centering
    \subfigure[Same-Version (Mutant)]{
        \includegraphics[scale=0.11]{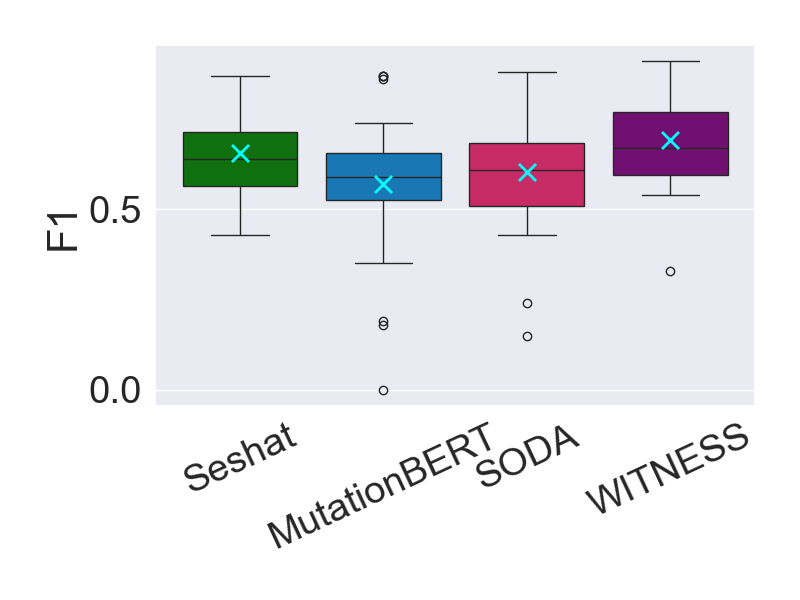}
    }
    \hfill
    \centering
    \subfigure[Cross-Version (Mutant)]{
        \includegraphics[scale=0.11]{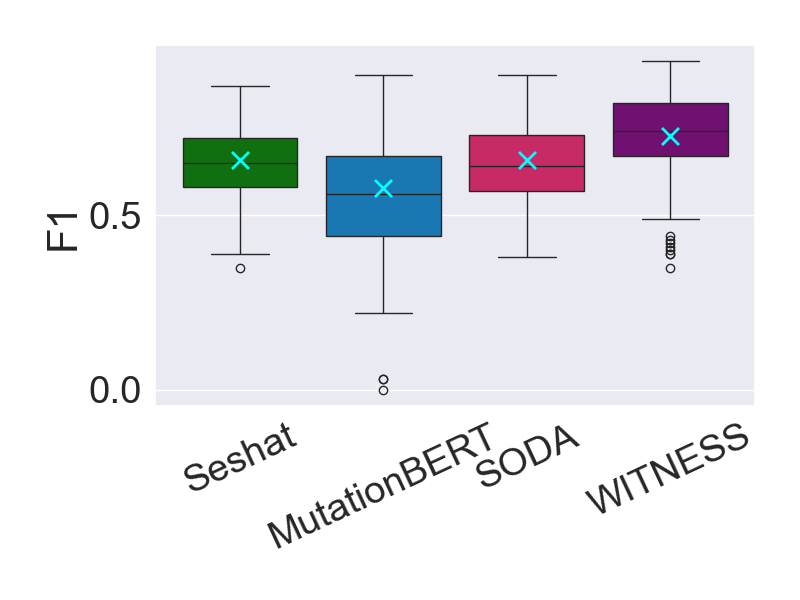}
    }
    \hfill
    \centering
    \subfigure[Cross-Project (Mutant)]{
        \includegraphics[scale=0.11]{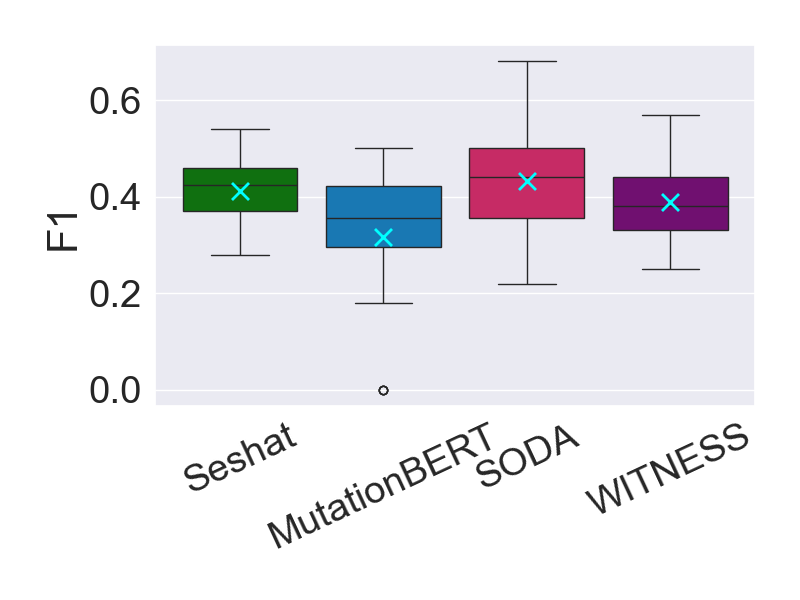}
    }
        \vfill
     \centering
    \subfigure[Same-Version (Mutation Score)]{
        \includegraphics[scale=0.11]{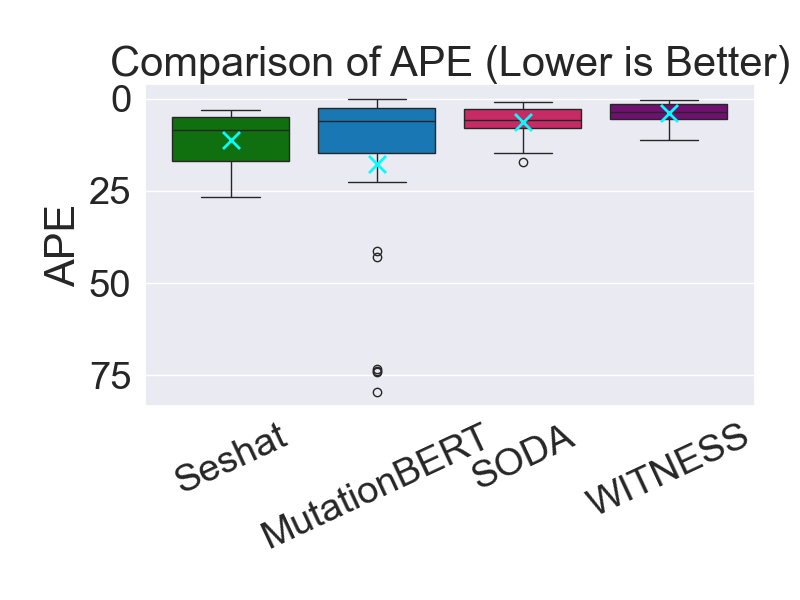}
    }
    \hfill
    \centering
    \subfigure[Cross-Version (Mutation Score)]{
        \includegraphics[scale=0.11]{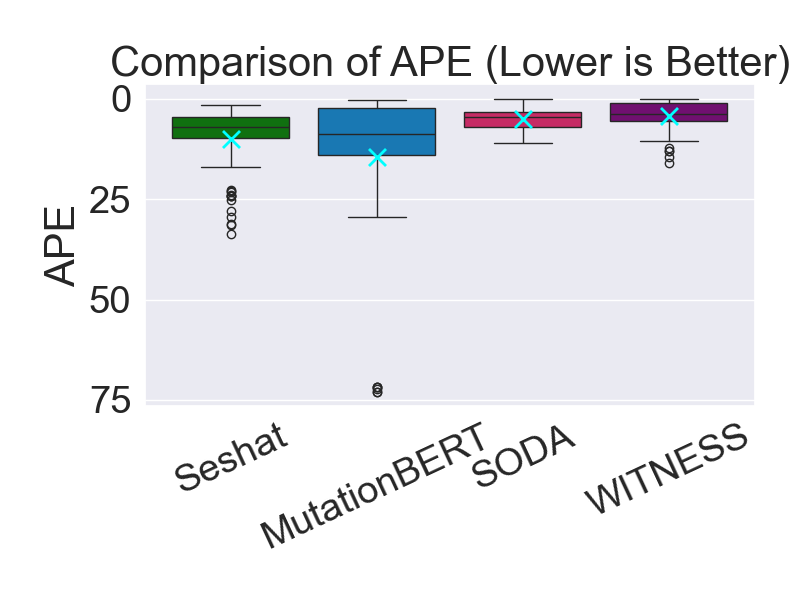}
    }
    \hfill
    \centering
    \subfigure[Cross-Project (Mutation Score)]{
        \includegraphics[scale=0.11]{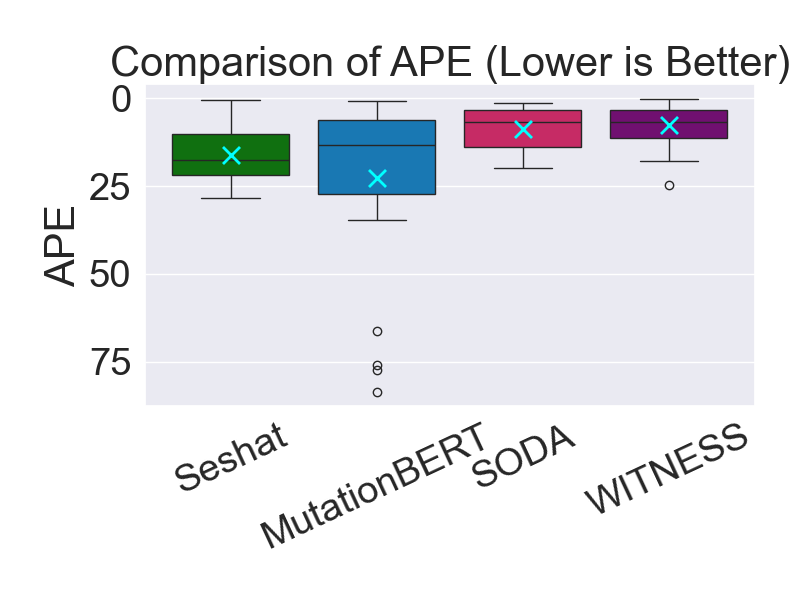}
    }
    \caption{Comparison of predictive performance using box plots. The first row shows the predictive performance in kill matrix prediction. The second row shows the performance in predicting whether a mutant is killed by a test suite. The third row presents the absolute prediction error between the predicted mutation score and the actual mutation score. The cross marker indicates the average value.}
    \label{fig:predictive_performance}
\end{figure*}

\textbf{\textit{Results.}} Table \ref{tab:performance_of_witness_with_baselines} presents the average predictive performance of WITNESS and the baselines. In Table \ref{tab:performance_of_witness_with_baselines}, the best results are shown in bold. Figs. \ref{fig:predictive_performance}(a)-(c) present boxplots of predictive performance, evaluated by F1-score, for Seshat, MutationBERT, SODA, and WITNESS in kill matrix prediction. Figs. \ref{fig:predictive_performance}(d)-(f) depict the predictive performance of the predictive approaches, measured by F1-score, in mutant killing prediction. Figs. \ref{fig:predictive_performance}(g)-(i) illustrate the outcomes of mutation score prediction. The cross-project scenario includes both one-to-one and many-to-one predictions. Each data point corresponds to the value of an evaluation metric for a single experiment.

Generally, WITNESS achieves better predictive performance in predicting the kill matrix and determining the killing results of each mutant compared to the baselines across all three scenarios. For WITNESS and the baselines, no single approach performs consistently well across all experiments and scenarios. This suggests that the insufficiency of MutationBERT and SODA stems in part from their evaluation being limited to only two experiments. The APE achieved by WITNESS is closer to 0 than that of the baselines, indicating that the predicted mutation score is closer to the actual mutation score across different scenarios. While under the cross-project scenario the mutant killing prediction of WITNESS is slightly lower than that of SODA, the APE of WITNESS is better than that of SODA. This is because the Random Forest model in WITNESS tends to predict more killing outcomes under the cross-project scenario.

As shown in Table \ref{tab:performance_of_witness_with_baselines} and Fig. \ref{fig:predictive_performance}, the predictive performance in the cross-project scenario across different approaches is lower than that in both the same-version and cross-version scenarios. This difference is due to the inherent variability and distinct characteristics of different projects, which also affect defect prediction in cross-project scenario \cite{khatri2022cross, qiu2019transfer, khatri2024predictive}. In the cross-version scenario, the four approaches achieve the highest average F1-scores compared to the same-version and cross-project scenarios. This is attributed to the fact that different versions within a project often contain many identical data elements.

\textit{Discussion of APE.} While APE evaluates predictive performance from a different perspective compared to F1-score in kill matrix prediction and mutant killing prediction, it should be considered alongside F1-score to demonstrate that a predictive approach can achieve a mutation score close to the actual value. APE is calculated based on mutant killing prediction. For each mutant, mutant killing prediction is a summary over all covered test cases. Even if the predictive performance on matrix prediction is low, a mutant is considered killed if just one of the covered test cases kills it, which can lead to better APE results. Specifically, the APE is small even when the predictive performance of kill matrix prediction is low, which may contradict the intuition that a low APE results from high predictive performance of kill matrix prediction. For example, under the same-version scenario for Csv\_5, the F1-scores of matrix prediction and mutant killing prediction, and the APE of Seshat are 0.75, 0.64, and 15.96, respectively. For MutationBERT, the corresponding results are 0.66, 0.54, and 2.13. While Seshat achieves higher predictive performance, its APE is much worse than that of MutationBERT.

Moreover, the APE does not reflect the lower prediction or higher prediction of the actual mutation score. The mutants that get killed typically make up the majority; thus, the actual mutation score is usually in the range of (50\%, 100\%]. Take an actual mutation score of 70\% as an example. For a predictive approach that tends to overpredict mutant killing, the maximum APE value is 30\% (the absolute value of 100\% minus 70\%). For a predictive approach that tends to underpredict the mutation score, the maximum APE value is 70\% (the absolute value of 0\% minus 70\%). Therefore, APE may favor approaches that tend to predict a higher mutation score than the actual mutation score. For the example of Csv\_5 under the same-version scenario, the actual mutation score is 71.28. The predicted mutation score of Seshat is 55.52, while that of MutationBERT is 73.40. In this case, MutationBERT predicts a higher mutation score than the actual one.

\begin{table}[tbp]
\centering
\scriptsize
\caption{Mean and Median of Both the Actual and Predicted Mutation Score}
\begin{tabular}{ccccccc}
\toprule
 \multirow{2.5}{*}{\textbf{Approach}} & \multicolumn{2}{c}{\textbf{Same-Version}} & \multicolumn{2}{c}{\textbf{Cross-Version}} & \multicolumn{2}{c}{\textbf{Cross-Project}} \\
\cmidrule(lr){2-3} \cmidrule(lr){4-5} \cmidrule(lr){6-7} 
  & \textbf{Mean} & \textbf{Median} & \textbf{Mean} & \textbf{Median} & \textbf{Mean} & \textbf{Median} \\
\midrule
Major      & 67.59 & 73.81 & 66.81 & 72.94 & 74.58 & 75.89 \\
\cmidrule(r){1-7} 
Seshat     & 56.67 & 55.77 & 56.80 & 59.02 & 58.50 & 59.17 \\
MutationBERT & 53.82 & 63.55 & 63.84 & 68.14 & 65.22 & 76.26 \\
SODA       & 73.13 & 77.89 & 71.40 & 79.89 & 71.59 & 72.10 \\
WITNESS    & 66.72 & 67.49 & 66.52 & 68.17 & 71.87 & 72.75 \\
\bottomrule
\end{tabular}
\label{tab:mutation_score}
\end{table}

Table \ref{tab:mutation_score} presents the mean and median of both the actual mutation scores and the predicted mutation scores from different predictive approaches. It is noteworthy that SODA typically predicts a mutation score higher than the actual mutation score, especially under the same-version and cross-version scenarios, which tends to lead to an overestimation of test suite effectiveness. Additionally, Seshat typically predicts a lower mutation score than the actual mutation score and generally yields the lowest predicted mutation score among all predictive approaches.

\subsubsection{Predictive Performance for All Mutant-Test Pairs and Those Where Mutations Occur Outside Source Methods}
~\

\textbf{\textit{Approach.}} The comparison with the baselines is based on a dataset that includes only mutant-test pairs where mutations occur inside source methods. Unlike the baselines, WITNESS is able to predict results for all mutant-test pairs, including those where mutations occur outside source methods. In this section, training and evaluation are conducted using datasets that include both inside-method and outside-method mutant-test pairs.

For datasets comprised of all mutant-test pairs, although the dataset splitting follows the same proportion as for datasets containing only inside-method mutant-test pairs (e.g., 80-10-10), the resulting training, validation, and test sets—which include mutant-test pairs where mutations occur outside source methods—differ from those based solely on inside-method pairs. Consequently, the trained models are different, and the predictive performance on datasets consisting of all mutant-test pairs typically differs from that on datasets containing only inside-method mutant-test pairs.

We present the predictive performance of WITNESS on datasets consisting of all mutant-test pairs. The predictive results for inside-method and outside-method mutant-test pairs are directly analyzed from the test sets, which include both types of pairs. Specifically, in each experiment, the predictive performance for all mutant-test pairs is evaluated on the entire test set. The predictive performance for inside-method and outside-method mutant-test pairs is then separately obtained from the respective subsets within the complete test set.

\begin{table*}[htbp]
\centering
\scriptsize
\caption{Average Predictive Performance of WITNESS}
\begin{threeparttable}
\begin{tabular}{cccccccccccccc}
\toprule
\multirow{2}{*}{} & \multicolumn{3}{c}{\textbf{All*}} & \multicolumn{3}{c}{\textbf{Inside-Method*}} & \multicolumn{3}{c}{\textbf{Outside-Method*}} & \multicolumn{3}{c}{\textbf{Inside-Method+}} & \\ 
 \cmidrule(lr){2-4}  \cmidrule(lr){5-7} \cmidrule(lr){8-10} \cmidrule(lr){11-13}
  & \textbf{SV} & \textbf{CV} & \textbf{CP} & \textbf{SV} & \textbf{CV} & \textbf{CP} & \textbf{SV} & \textbf{CV} & \textbf{CP} & \textbf{SV} & \textbf{CV} & \textbf{CP} & \\
\midrule
& 0.67 & 0.74 & 0.46 & 0.67 & 0.73 & 0.46 & 0.65 & 0.84 & 0.56 & 0.67 & 0.73 & 0.46 & Precision \\
Kill Matrix & 0.78 & 0.87 & 0.65 & 0.78 & 0.87 & 0.66 & 0.69 & 0.88 & 0.58 & 0.79 & 0.87 & 0.66 & Recall \\                  
& 0.72 & 0.80 & 0.53 & 0.72 & 0.79 & 0.53 & 0.64 & 0.86 & 0.51 & 0.72 & 0.79 & 0.53 & F1-score \\
\midrule
& 0.69 & 0.73 & 0.38 & 0.69 & 0.74 & 0.38 & 0.47 & 0.50 & 0.28 & 0.70 & 0.74 & 0.38 & Precision \\
Mutant Killing & 0.70 & 0.73 & 0.43 & 0.70 & 0.73 & 0.43 & 0.69 & 0.86 & 0.53 & 0.70 & 0.73 & 0.42 & Recall \\
& 0.69 & 0.72 & 0.39 & 0.69 & 0.73 & 0.39 & 0.54 & 0.59 & 0.31 & 0.69 & 0.73 & 0.39 & F1-score \\
\midrule
Mutation Score & 4.07 & 4.15 & 7.84 & 4.06 & 4.18 & 7.92 & 20.51 & 22.68 & 27.69 & 3.94 & 4.22 & 7.73 & APE \\
\bottomrule
\end{tabular}
    \begin{tablenotes}
	\item SV, CV, and CP represent Same-Version, Cross-Version, and Cross-Project, respectively.
    \end{tablenotes}
\end{threeparttable}
\label{tab:predictive_performance_all}
\end{table*}

\textbf{\textit{Results.}} As shown in Table \ref{tab:predictive_performance_all}, for models trained on all mutant-test pairs, we report predictive performance on all mutant-test pairs (All*), on inside-method mutant-test pairs (Inside-Method*), and on outside-method mutant-test pairs (Outside-Method*).

For models trained on all mutant-test pairs, the predictive performance on all mutant-test pairs (All*) is similar to that on inside-method mutant-test pairs (Inside-Method*). However, the predictive performance on outside-method mutants is lower than that on inside-method mutants. One reason is that there may be a much smaller number of mutant-test pairs involving outside-method mutants, meaning incorrect predictions on a few of them can significantly reduce overall performance. Another reason is that outside-method mutants typically involve the declaration of member variables, which can be of user-defined types. In such cases, the mutation may affect only a parameter of a constructor in the user-defined class. Without full information about the user-defined class, it may be difficult to make accurate predictions.

For dataset splitting and model training conducted only on inside-method mutant-test pairs, we include the results achieved by WITNESS (Inside-Method+) for reference. Although the dataset and trained models differ, the predictive performance on Inside-Method* and Inside-Method+ is similar. Additionally, the performance on All* and Inside-Method+ is also comparable.

\begin{tcolorbox}[colback=black!5!white, colframe=black!75!black]
\textbf{Answer for RQ1: WITNESS generally achieves higher effectiveness in predictive performance compared to baselines across various scenarios. WITNESS can predict outcomes for all mutant-test pairs, including those outside source methods, and its predictive performance is comparable to that achieved on inside-method pairs alone.} 
\end{tcolorbox}

\subsection{RQ2: Predictive Performance for Different Mutant Killing Reasons}

\textbf{\textit{Approach.}} We compare the prediction results for test cases that kill mutants due to three reasons: assertion failure (FAIL), timeout (TIME), and exception (EXC).

\begin{table*}[h]
\centering
\scriptsize
\caption{Average Predictive Performance Under Different Killing Reasons for WITNESS and the Baselines}
\begin{threeparttable}
\begin{tabular}{ccccccccccccc}
\toprule
\multirow{2}{*}{} & \multicolumn{3}{c}{\textbf{Seshat}} & \multicolumn{3}{c}{\textbf{MutationBERT}} & \multicolumn{3}{c}{\textbf{SODA}} & \multicolumn{3}{c}{\textbf{WITNESS}} \\ 
 \cmidrule(lr){2-4}  \cmidrule(lr){5-7}  \cmidrule(lr){8-10} \cmidrule(lr){11-13}
  & \textbf{SV} & \textbf{CV} & \textbf{CP} & \textbf{SV} & \textbf{CV} & \textbf{CP} & \textbf{SV} & \textbf{CV} & \textbf{CP} & \textbf{SV} & \textbf{CV} & \textbf{CP} \\
\midrule
FAIL & 0.70 & 0.79 & 0.54 & 0.58 & 0.77 & 0.62 & 0.76 & 0.86 & 0.62 & \textbf{0.78} & \textbf{0.87} & \textbf{0.69} \\ 
\midrule
TIME & 0.63 & 0.71 & 0.59 & 0.53 & 0.66 & 0.70 & 0.63 & 0.71 & 0.68 & \textbf{0.71} & \textbf{0.78} & \textbf{0.72} \\ 
\midrule
EXC & 0.71 & 0.77 & 0.52 & 0.58 & 0.73 & 0.59 & 0.77 & 0.81 & 0.58 & \textbf{0.80} & \textbf{0.86} & \textbf{0.65} \\ 
\bottomrule
\end{tabular}
    \begin{tablenotes}
	\item SV, CV, and CP represent Same-Version, Cross-Version, and Cross-Project, respectively.
    \end{tablenotes}
\end{threeparttable}
\label{tab:killing_reasons}
\end{table*}

\begin{figure*}[htbp]
    \centering
    \subfigure[Same-Version - FAIL]{
        \includegraphics[scale=0.11]{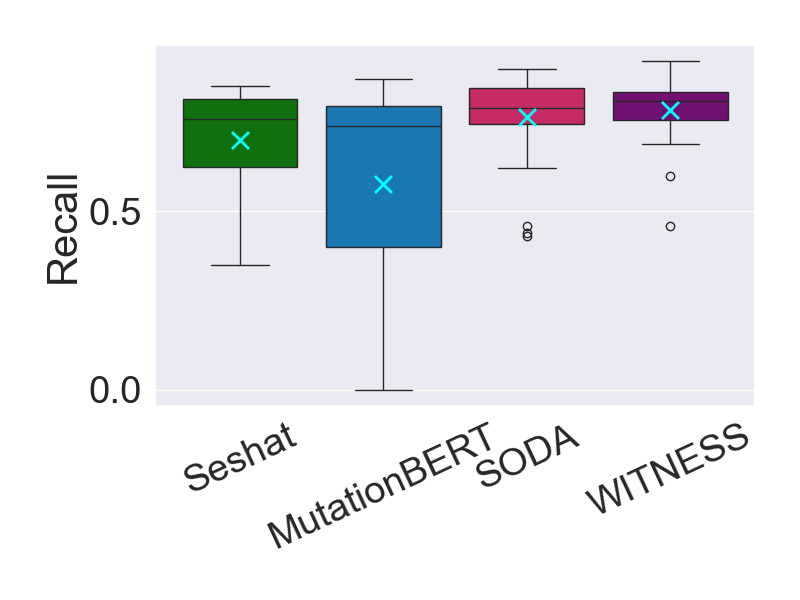}
    }
    \hfill
    \centering
    \subfigure[Cross-Version - FAIL]{
        \includegraphics[scale=0.11]{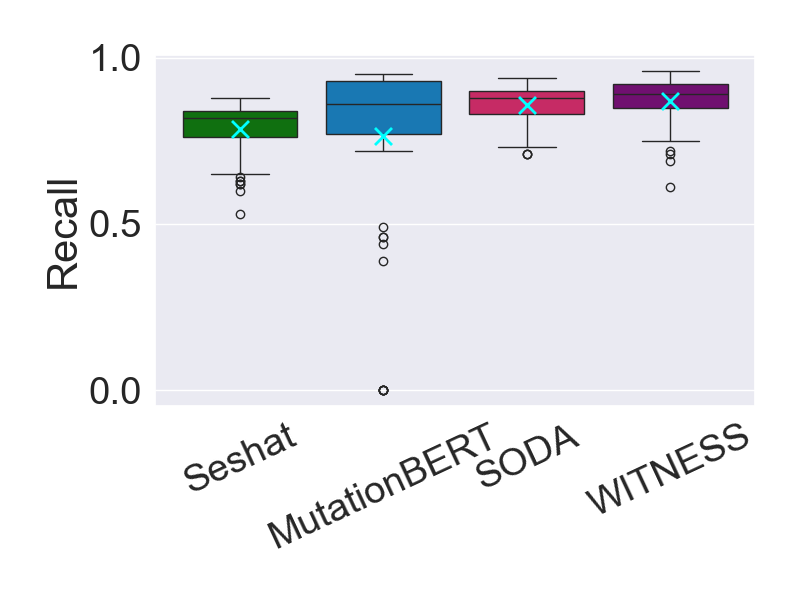}
    }
    \hfill
    \centering
    \subfigure[Cross-Project - FAIL]{
        \includegraphics[scale=0.11]{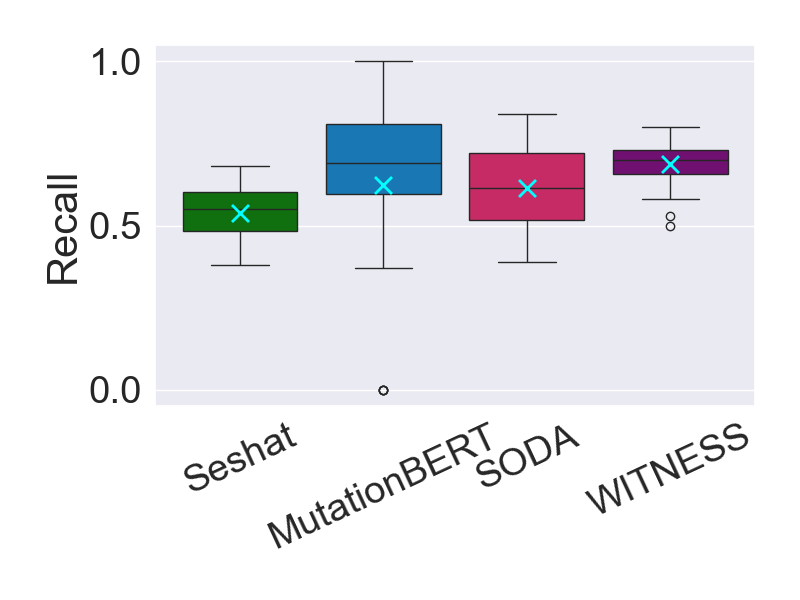}
    }
    \vfill
    \centering
    \subfigure[Same-Version - TIME]{
        \includegraphics[scale=0.11]{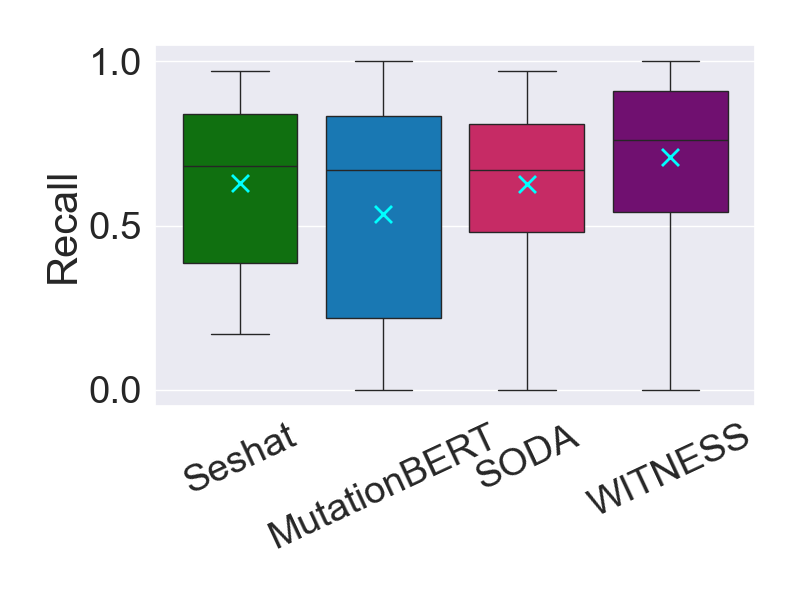}
    }
    \hfill
    \centering
    \subfigure[Cross-Version - TIME]{
        \includegraphics[scale=0.11]{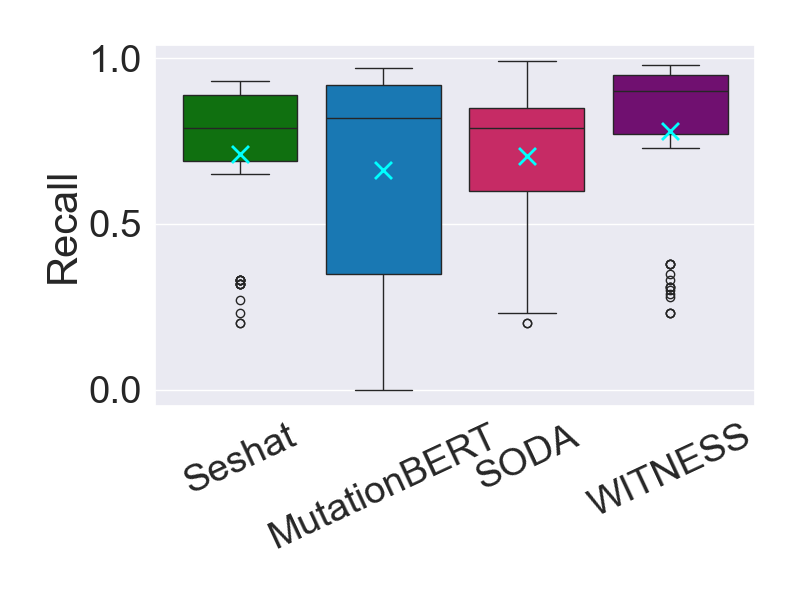}
    }
    \hfill
    \centering
    \subfigure[Cross-Project - TIME]{
        \includegraphics[scale=0.11]{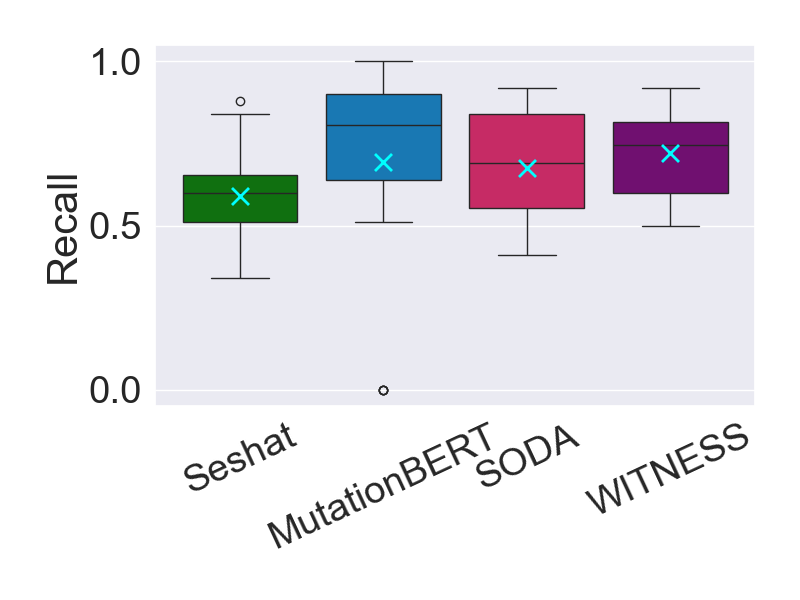}
    }
    \vfill
    \centering
    \subfigure[Same-Version - EXC]{
        \includegraphics[scale=0.11]{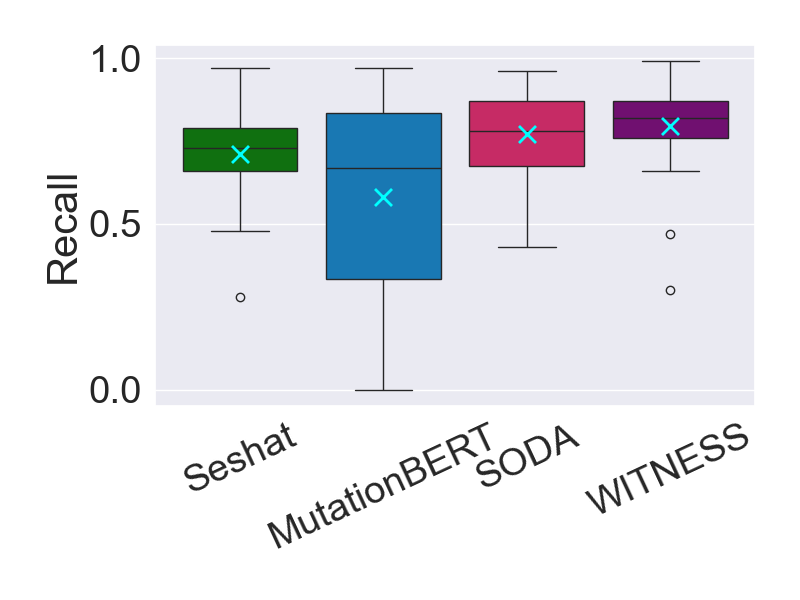}
    }
    \hfill
    \centering
    \subfigure[Cross-Version - EXC]{
        \includegraphics[scale=0.11]{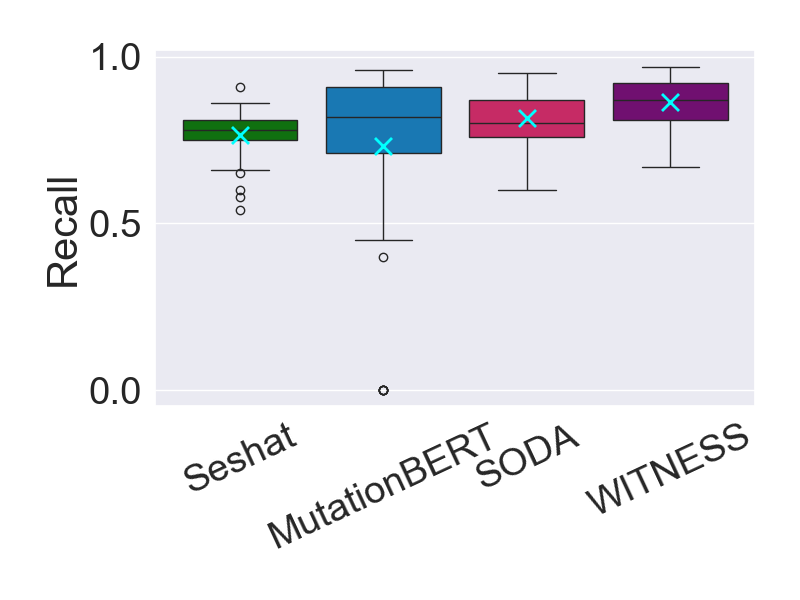}
    }
    \hfill
    \centering
    \subfigure[Cross-Project - EXC]{
        \includegraphics[scale=0.11]{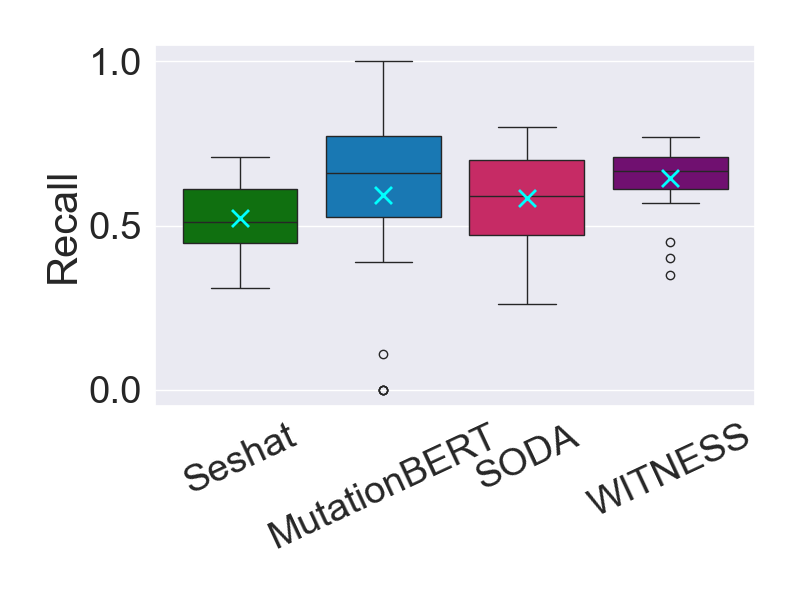}
    }
    \caption{Comparison of predictive performance using box plots under different killing reasons across various predictive approaches. ‘FAIL,’ ‘TIME,’ and ‘EXC’ represent test case failures due to assertion failure, timeout, and exception, respectively. The cross marker indicates the average value.}
    \label{fig:killing_reasons}
\end{figure*}

\textbf{\textit{Results.}} Table \ref{tab:killing_reasons} and Fig. \ref{fig:killing_reasons} presents a comparison of predictive performance under three killing reasons across different predictive approaches. In Table \ref{tab:killing_reasons}, the best results are shown in bold. Figs. \ref{fig:killing_reasons}(a) through \ref{fig:killing_reasons}(c) compare predictive performance for test case failures due to assertion failures. Figs. \ref{fig:killing_reasons}(d) through \ref{fig:killing_reasons}(f) and Figs. \ref{fig:killing_reasons}(g) through \ref{fig:killing_reasons}(i) compare predictive performance for test case failures due to timeouts and exceptions, respectively. 

As shown in Table \ref{tab:killing_reasons} and Fig. \ref{fig:killing_reasons}, WITNESS generally achieves higher predictive performance across different killing reasons. In the cross-project scenario, MutationBERT appears to perform better than the other approaches, as shown in Fig. \ref{fig:killing_reasons}. This is because MutationBERT’s recall values range from 0.0 to 1.0, resulting from its predicted kill matrices consisting entirely of zeros (no test cases kill mutants) or entirely of ones (all test cases kill mutants) in several experiments. This highlights the highly volatile predictive performance of MutationBERT. In contrast, WITNESS and the other baselines, Seshat and SODA, do not produce a recall as low as 0.0 under the cross-project scenario.

Given that the main purpose of mutation testing is to evaluate test suite effectiveness, which closely relates to assertion failures, Figs. \ref{fig:killing_reasons}(a) through \ref{fig:killing_reasons}(c) demonstrate that WITNESS is much more suitable for evaluating test suite effectiveness than the baselines.

\begin{tcolorbox}[colback=black!5!white, colframe=black!75!black]
\textbf{Answer for RQ2: WITNESS generally achieves higher predictive performance across different killing reasons. Since mutation testing primarily aims to evaluate test suite effectiveness, which is closely tied to assertion failures, WITNESS is particularly well-suited for this purpose.}
\end{tcolorbox}

\subsection{RQ3: Prediction of Outcomes for Fewer-Covered Mutants}

\textbf{\textit{Approach.}} Fewer-covered mutants are those that are covered by only a small number of test cases. For example, among all test cases that could cover mutants, some mutants are covered by just one or two test cases.  For fewer-covered mutants, since a smaller number of mutant-test pairs are formed, incorrect predictions for these pairs can easily lead to inaccurate predictions for fewer-covered mutants. This makes comparing the predictive performance for fewer-covered mutants particularly meaningful. Additionally, analyzing the predictive results for fewer-covered mutants helps evaluate the predictive capability of different predictive approaches on sparse data.

We compare the predictive performance for mutants covered by fewer test cases. We focus on mutants with less than or equal to 2\% of test cases covering them among all test cases that could cover mutants. We choose 2\% because, in some smaller-sized test sets under the same-version scenario, there are no mutants covered by less than or equal to 1\% of the test cases.

\begin{figure*}[htbp]
    \centering
    \subfigure[Same-Version]{
        \includegraphics[scale=0.11]{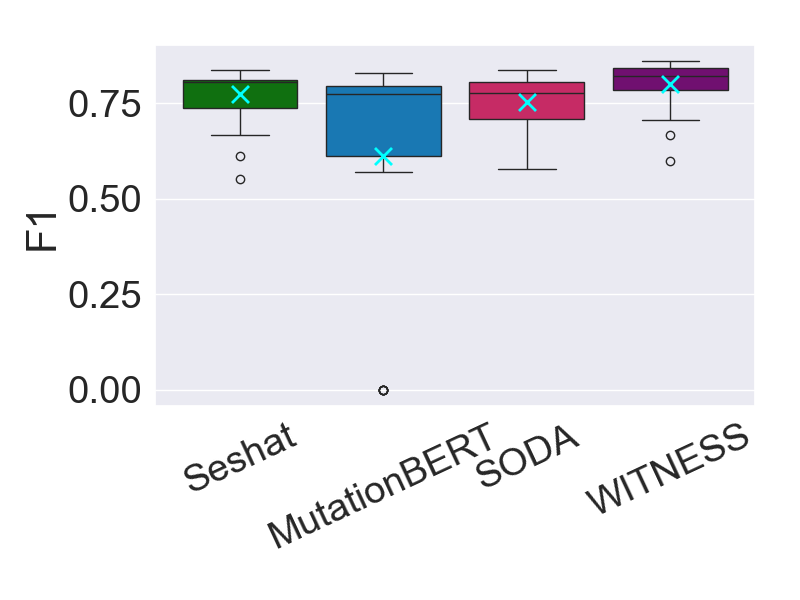}
    }
    \hfill
    \centering
    \subfigure[Cross-Version]{
        \includegraphics[scale=0.11]{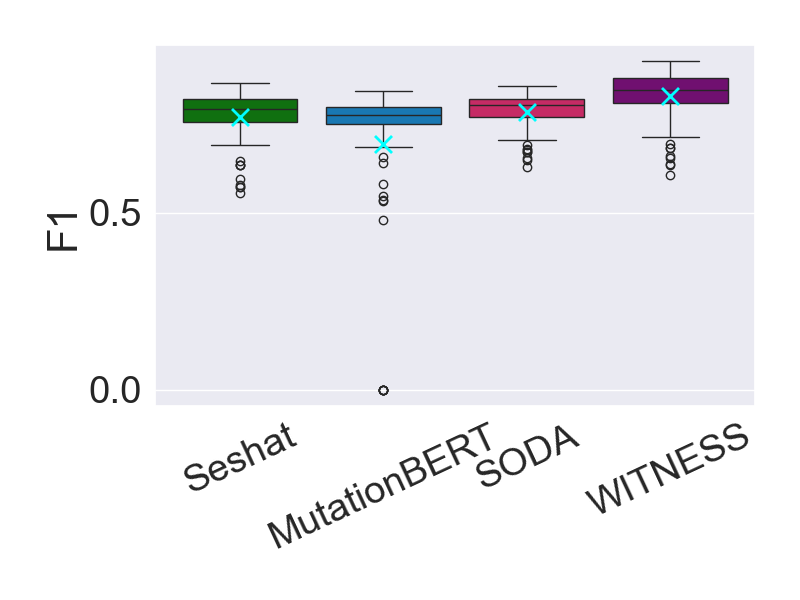}
    }
    \hfill
    \centering
    \subfigure[Cross-Project]{
        \includegraphics[scale=0.11]{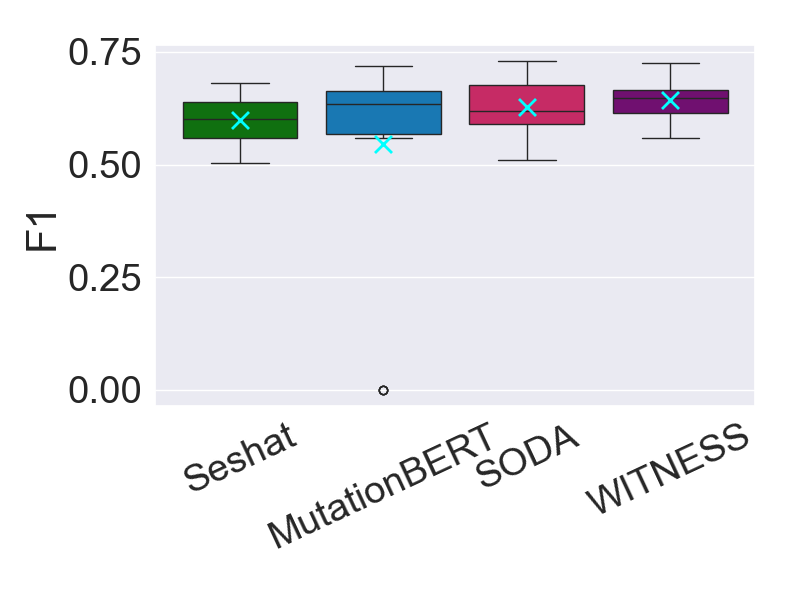}
    }
    \caption{Comparison of F1-scores in predicting the results of fewer-covered mutants using box plots under different scenarios. The cross marker indicates the average value.}
    \label{fig:fewer_covered}
\end{figure*}

\begin{table}[tbp]
\centering
\scriptsize
    \caption{Average F1-score in Predicting the Results of Mutants Covered by Less Than or Equal to 2\% of the Test Cases Among All Test Cases That Could Cover Mutants Across Different Scenarios}
    \begin{tabular}{cccccc}
    \toprule
    & \textbf{Seshat} & \textbf{MutationBERT} & \textbf{SODA} & \textbf{WITNESS} \\
    \midrule
    Same-Version & 0.77 & 0.61 & 0.75 & \textbf{0.80} \\
    Cross-Version & 0.77 & 0.70 & 0.79 & \textbf{0.83} \\
    Cross-Project & 0.60 & 0.55 & 0.63 & \textbf{0.64} \\
    \cmidrule(lr){1-5} 
    \textit{Avg.}    & 0.71 & 0.62 & 0.72 & \textbf{0.76} \\
    \bottomrule
    \end{tabular}
    \label{tab:fewer_covered}
\end{table}

\textbf{\textit{Results.}} Table \ref{tab:fewer_covered} presents the average F1-scores for predicting the results of fewer-covered mutants, achieved by all four predictive approaches. The best values in Table \ref{tab:fewer_covered} are highlighted in bold. Fig. \ref{fig:fewer_covered} presents the corresponding boxplots. Overall, WITNESS demonstrates higher performance in predicting the outcomes of fewer-covered mutants compared to the baselines.

\begin{tcolorbox}[colback=black!5!white, colframe=black!75!black]
\textbf{Answer for RQ3: WITNESS generally achieves higher predictive performance in predicting the results of fewer-covered mutants compared to the baselines.} 
\end{tcolorbox}

\subsection{RQ4: Contribution of Features}

\textbf{\textit{Approach.}} Knowing how different features affect model predictions helps in interpreting the model’s behavior. Feature importance quantifies the contribution of each individual feature to a machine learning model’s predictions. This helps in understanding which features contribute most to the predictions. We analyze feature importance to gauge the impact strength of each feature used in WITNESS. We include all mutant-test pairs in this analysis, including those where mutations occur outside source methods, as there is no need to perform comparisons with baselines.

We analyze feature importance under the same-version, cross-version, and cross-project scenarios. However, the calculation of feature importance does not involve the test sets. Under the same-version scenario, the dataset is split 80-10-10. Under the cross-version scenario, the dataset is split with 90\% for the training set and 10\% for the validation set in earlier versions, with the test set comprising all data from later versions. If the earlier version is fixed, the change of the test set does not affect the feature importance. Thus, we analyze feature importance for the 31 versions used across 6 projects. Under the cross-project scenario, we analyze feature importance in many-to-one prediction, as the training and validation sets are composed of data from different projects.

After obtaining the raw feature importance from models trained on different datasets, we apply min-max normalization to standardize the values. We then aggregate all normalized feature importances within each scenario and calculate the average value to determine the final feature importance for that specific scenario. For the Random Forest and LightGBM models, we separately analyze the importance of the various features used in WITNESS.

\begin{figure}[htbp]
    \centering
    \subfigure[Feature importance from random forest model.]{
        \includegraphics[scale=0.15]{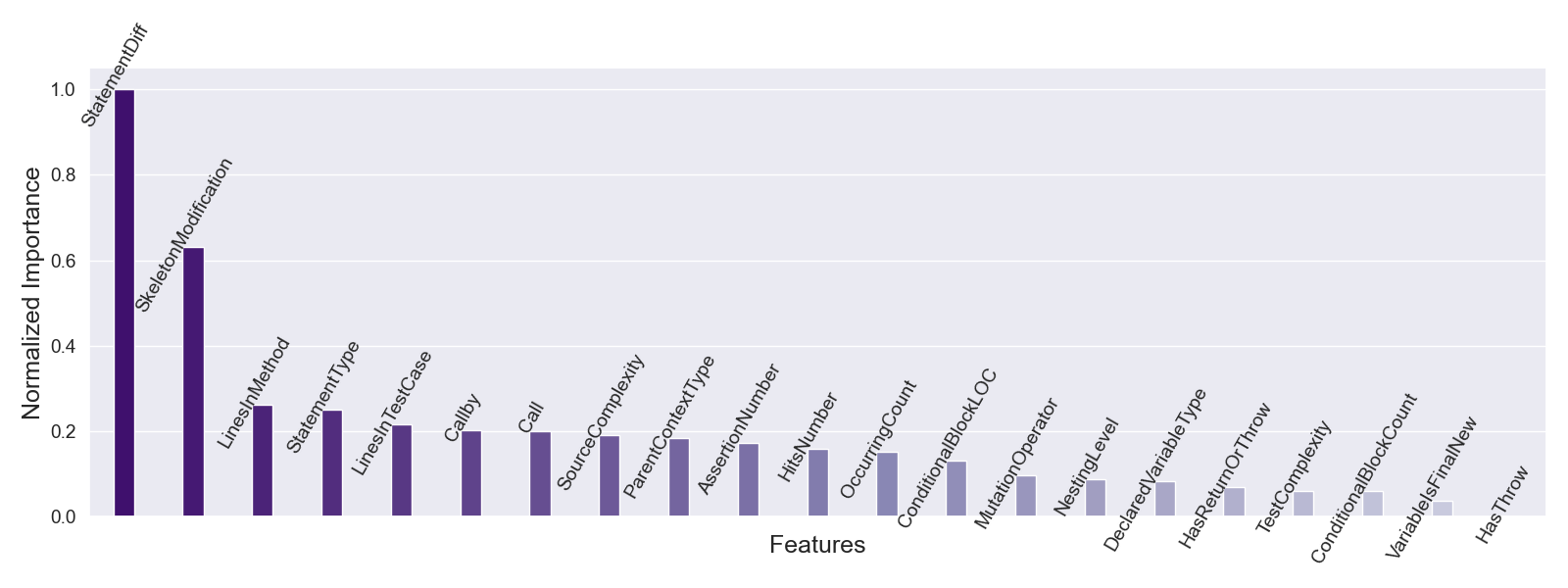}
    }
    \vfill
    \centering
    \subfigure[Feature importance from LightGBM model.]{
        \includegraphics[scale=0.15]{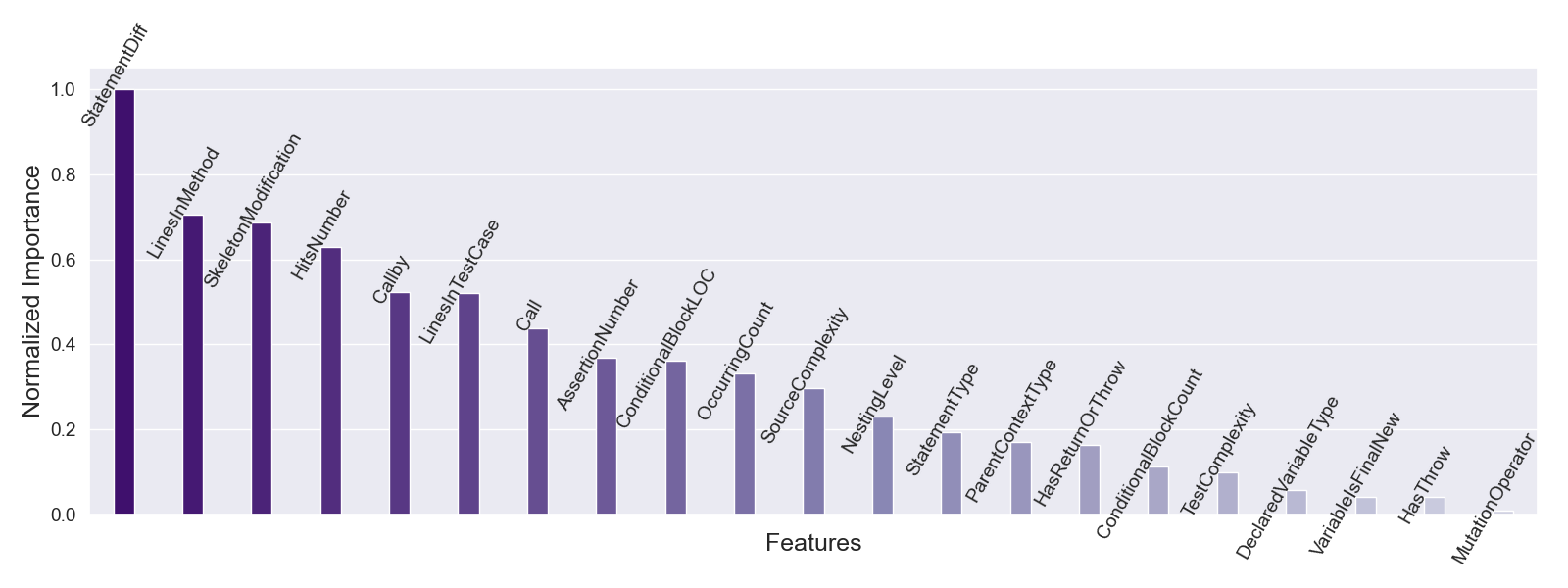}
    }
    \caption{Feature importance from the two adopted machine learning models under the same-version scenario.}
    \label{fig:feature_importance_same_version}
\end{figure}

\begin{figure}[htbp]
    \centering
    \subfigure[Feature importance from random forest model.]{
        \includegraphics[scale=0.15]{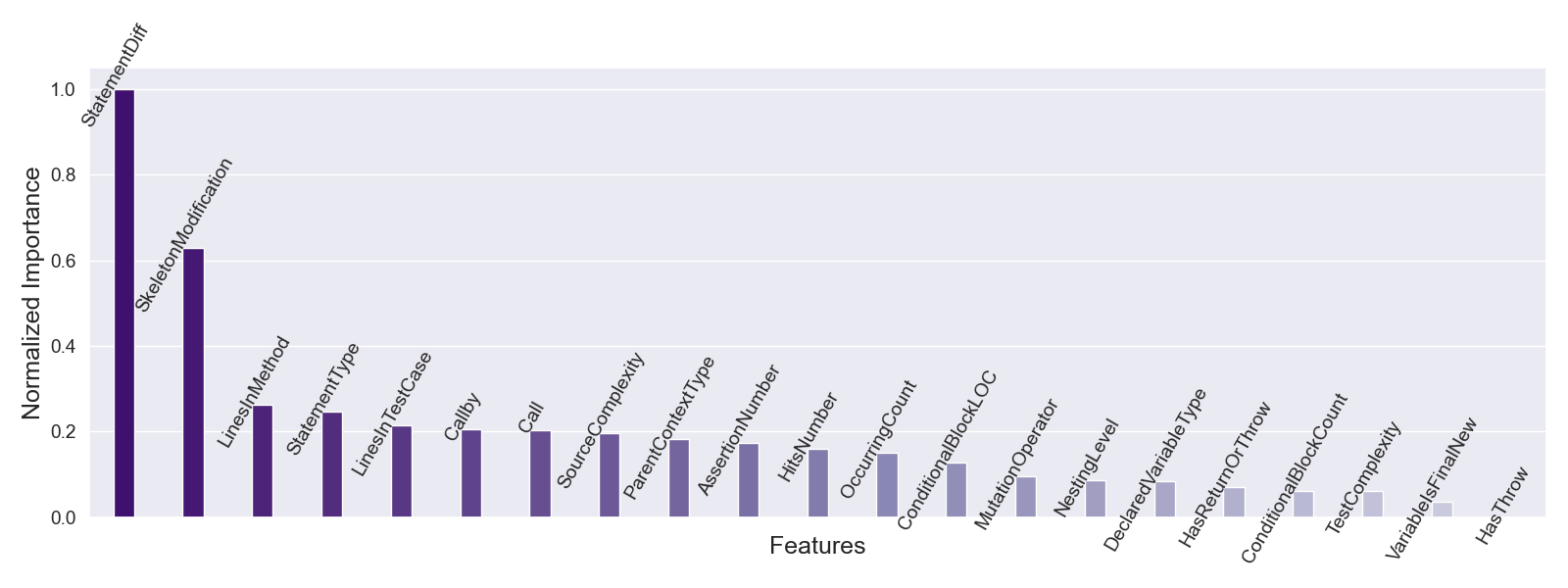}
    }
    \vfill
    \centering
    \subfigure[Feature importance from LightGBM model.]{
        \includegraphics[scale=0.15]{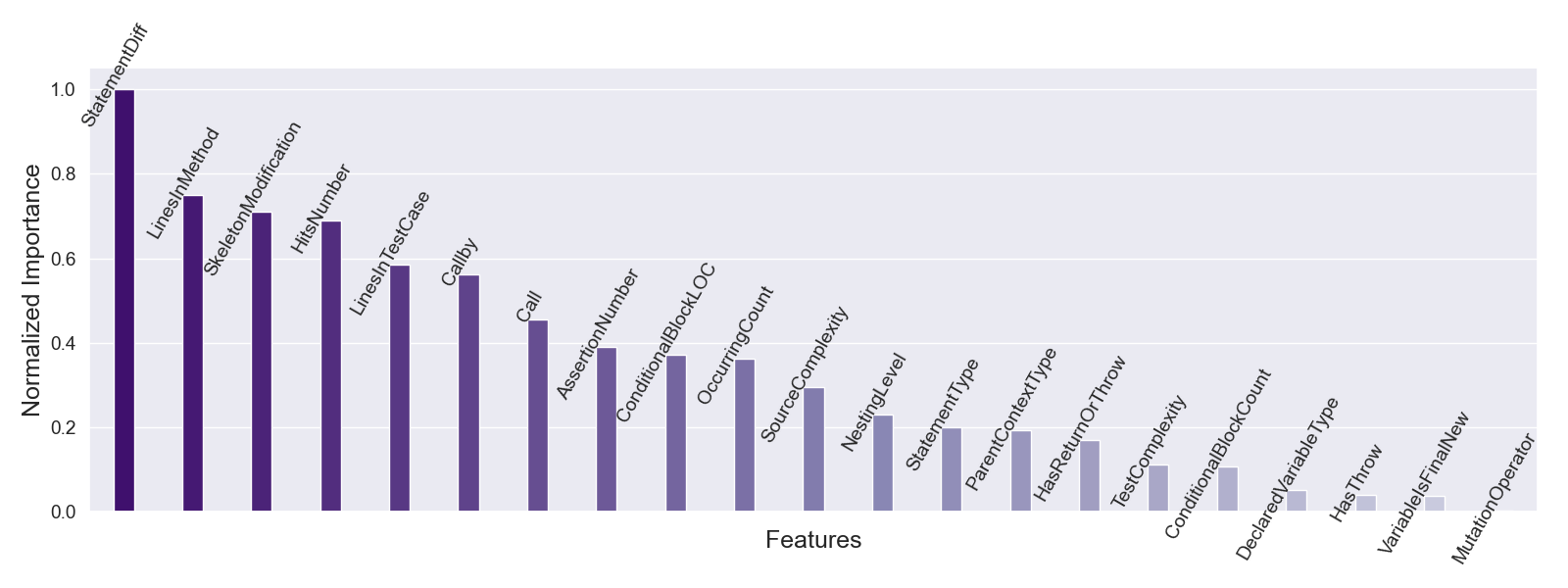}
    }
    \caption{Feature importance from the two adopted machine learning models across 31 project versions, with the dataset split 90\% for the training set and 10\% for the validation set, respectively.}
    \label{fig:feature_importance_31}
\end{figure}

\begin{figure}[htbp]
    \centering
    \subfigure[Feature importance from random forest model.]{
        \includegraphics[scale=0.15]{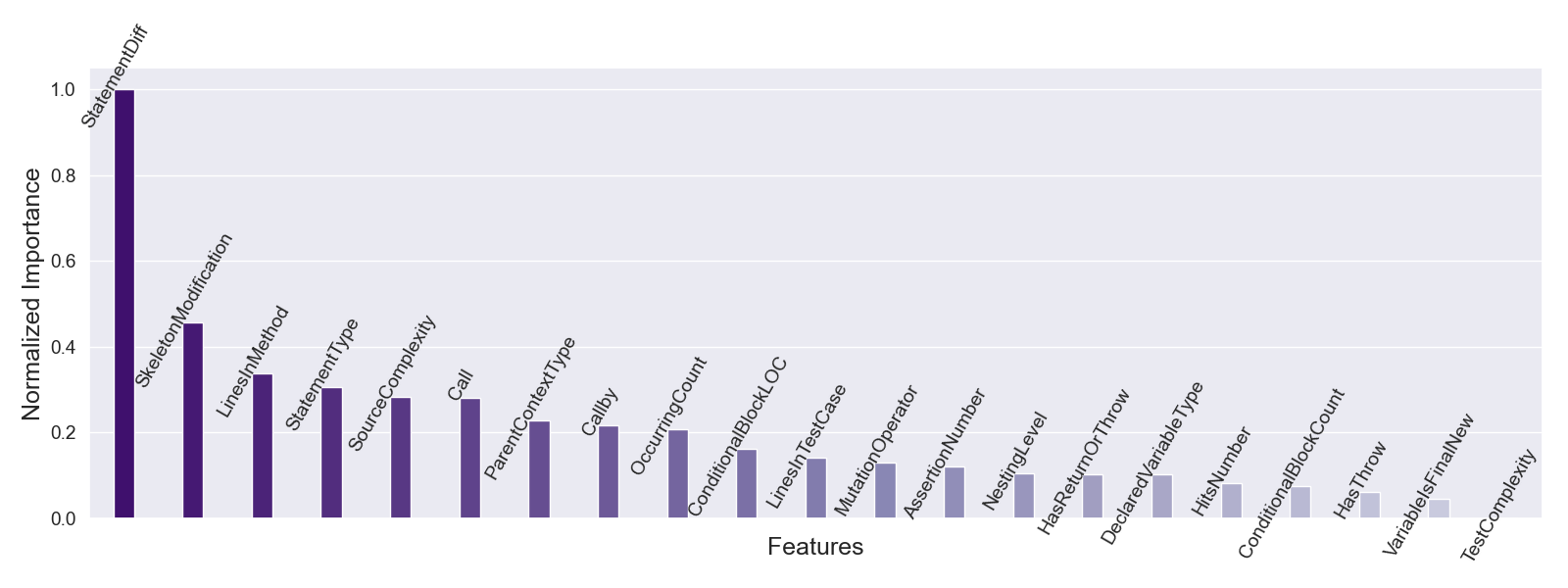}
    }
    \vfill
    \centering
    \subfigure[Feature importance from LightGBM model.]{
        \includegraphics[scale=0.15]{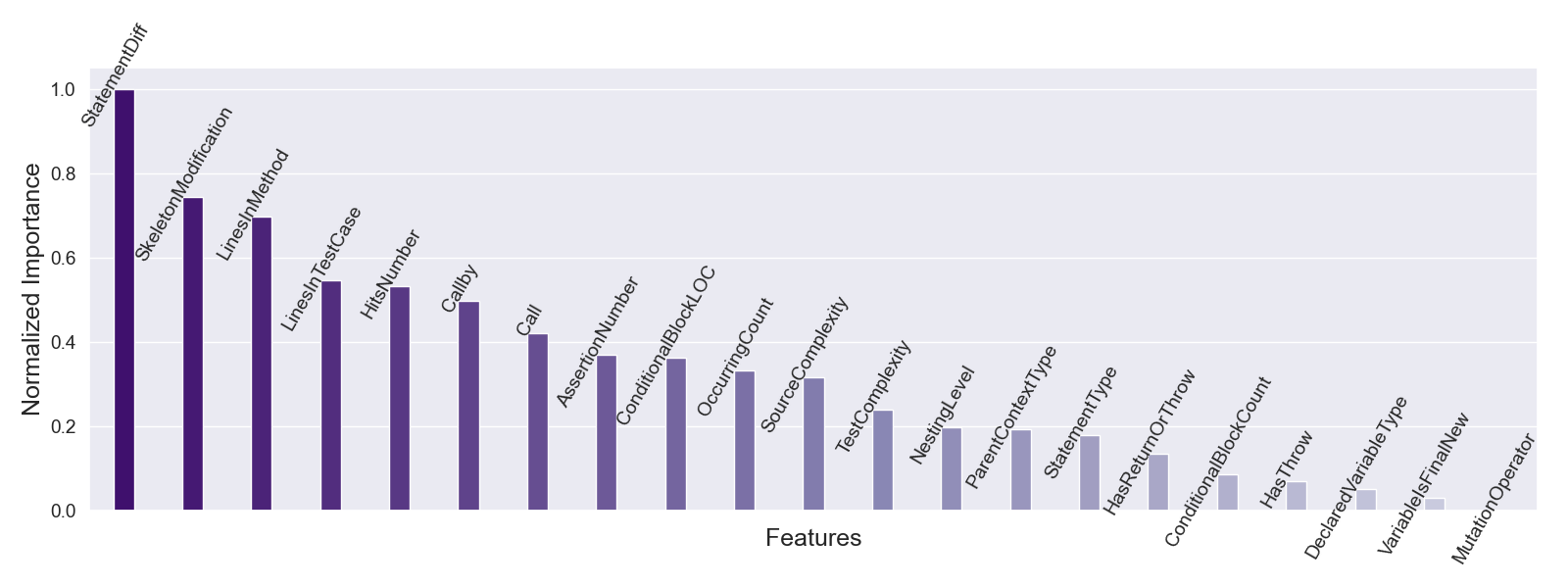}
    }
    \caption{Feature importance from the two adopted machine learning models under the cross-project scenario. }
    \label{fig:feature_importance_cross_project}
\end{figure}

\textbf{\textit{Results.}} Figs. \ref{fig:feature_importance_same_version}, \ref{fig:feature_importance_31}, and \ref{fig:feature_importance_cross_project} show the feature importance under the same-version scenario, across 31 project versions with the dataset split by 90\% and 10\%, and under the cross-project scenario, respectively. Sub-figure (a) of the three figures displays the feature importance obtained from the Random Forest model, while sub-figure (b) displays the feature importance from the LightGBM model. In all three figures, the features are sorted in descending order of importance.

Clearly, the two features, \textit{StatementDiff} and \textit{SkeletonModification}, which belong to the source code change category, contribute most to model performance, especially \textit{StatementDiff}. The feature \textit{StatementDiff} ranks first in all experimental settings. The feature \textit{SkeletonModification} ranks second in the Random Forest model and third in the LightGBM model under both the same-version and cross-version scenarios. Under the cross-project scenario, \textit{SkeletonModification} ranks second in both the Random Forest and LightGBM models.

The two features, \textit{HasThrow} and \textit{MutationOperator}, are among the lowest-ranked features in terms of importance. However, \textit{HasThrow} shows importance under the cross-project scenario, and \textit{MutationOperator} plays a non-negligible role in the Random Forest model. As for \textit{HitsNumber}, the only dynamic feature used in WITNESS, it ranks relatively higher in the LightGBM model than in the Random Forest model, indicating its notable contribution in determining whether a test case will kill a mutant.

Given the same source code, there are numerous ways to perform mutations, leading to the generation of various mutants. For a test case that executes the source code and forms various mutant-test pairs, the differentiation of these pairs largely depends on the specific ways the source code is mutated. This variation in mutation may result in the two features, \textit{StatementDiff} and \textit{SkeletonModification}, which include information before and after the mutation, contributing the most to model performance. 

It is noteworthy that all baseline approaches, Seshat, MutationBERT, and SODA, include the feature \textit{StatementDiff}. Among the features used by Seshat, two notable ones are ‘\textit{before}’ and ‘\textit{after}’, which refer to the changed parts of the statement before and after mutation, respectively. For the features used by MutationBERT and SODA, the corresponding features are ‘\textit{new\_line}’, ‘\textit{before}’, and ‘\textit{after}’. The feature ‘\textit{new\_line}’ represents the entire statement after mutation, while the features ‘\textit{before}’ and ‘\textit{after}’ correspond to the same two features used in Seshat. Based on the observation of the most important features in WITNESS, an actionable insight is that fine-grained predictive mutation testing approaches should prioritize including features related to source code changes.

\begin{tcolorbox}[colback=black!5!white, colframe=black!75!black]
\textbf{Answer for RQ4:  The features \textit{StatementDiff} and \textit{SkeletonModification}, which include information before and after the mutation, are the most important among the features adopted in WITNESS, especially \textit{StatementDiff}.} 
\end{tcolorbox}

\subsection{RQ5: Efficiency}

\textbf{\textit{Approach.}} Intuitively, WITNESS achieves significantly higher efficiency in predicting the kill matrix compared to the three baselines, as it utilizes simple machine learning models. We compare the efficiency of four predictive approaches across different scenarios. In the cross-version scenario, we use the earliest version of a project to predict subsequent versions. This choice is motivated by one of the baselines, Seshat, which specifically evaluates efficiency in this context. The time reported by Seshat serves as a reference for the time measured on our machine. Additionally, in the cross-project scenario, we focus on many-to-one prediction, as it involves training on different projects.

We focus on the prediction time, as it is commonly reported by the baselines, and Zhao et al. \cite{zhao2024spotting} specifically report the prediction time for SODA. To align with baselines, the prediction time refers to the duration needed to predict results for the mutant-test pairs where mutations occur within source methods. It is noteworthy that while the training of predictive models can be done offline, it requires significant time for predictive approaches such as MutationBERT and SODA. Take training four projects - JacksonCore, Gson, Cli, and Csv - to predict project Lang as an example. When measured on our server, the training of SODA on two V100 GPUs takes more than 5 days (7,270 minutes and 53.42 seconds), whereas WITNESS completes it in just 61.64 seconds on a CPU. The training time of SODA, which is on the scale of days, takes significantly longer than directly conducting mutation testing, hindering the practical application of SODA.

To measure the time taken, we run the inference program without any other programs running on a server. For each prediction, we repeat the process 10 times to account for variability in prediction time when predicting a kill matrix, taking the average value as the final prediction time. We measure the prediction times of Seshat, MutationBERT, and SODA on GPUs. In contrast, WITNESS’s prediction time is measured on a CPU. If WITNESS still achieves a substantially higher speed-up in prediction time compared to the baselines, this underscores its greater efficiency in predicting the kill matrix. 

As pointed out in Section 4.4, we assess whether each mutant is predicted to be killed by a test suite based on the predicted kill matrix. We do not train separate models for kill matrix prediction and mutant killing results. In MutationBERT and SODA, the test set is split based on every 10,000 mutant-test pairs. For example, when predicting Csv\_5 with 22,968 mutant-test pairs, the test set is split into three files — two containing 10,000 pairs each and one with 2,968 pairs. This setup enables faster predictions through parallel predicting. Although splitting the test set could result in faster prediction speeds compared to using an unsplit test set, we avoid splitting the test set as MutationBERT and SODA do. The reason is that the datasets constructed by MutationBERT and SODA do not include identifiers for individual test cases, making it impossible to trace each prediction back to the correct mutant-test pair. As a result, accurate mutant killing results cannot be obtained from the predicted outcomes of mutant-test pairs for MutationBERT and SODA. Moreover, for MutationBERT and SODA, even without splitting the test set, running on GPUs still provides acceleration benefits. For example, when using Csv\_1 to predict Csv\_15, MutationBERT takes 30,818 seconds on a CPU but only 1,191 seconds on a single GPU on our server, achieving an acceleration of approximately 25.88 times.

We present the time required for actual mutation testing, which involves executing the test suite against each generated mutant. We compare the prediction time of the four predictive approaches with the time taken for actual mutation testing. If the prediction time of a predictive approach exceeds the time required for actual mutation testing, it will hinder the practical utility of that approach. 

\begin{table*}[htbp]
\centering
\scriptsize
\caption{Time Taken by Predictive Approaches and Speedup Achieved by WITNESS under the Same-Version Scenario}
\adjustbox{width=1.0\columnwidth}{%
\begin{threeparttable}
\begin{tabular}{@{}ccccccccccc@{}} 
\toprule
\multirow{2.5}{*}{\textbf{Project ID}} & \multirow{2.5}{*}{\textbf{Version}} & \textbf{\# Mutant-Test} & \multicolumn{4}{c}{\textbf{Prediction Approaches}} & \multicolumn{3}{c}{\textbf{WITNESS Speed Up}} \\
\cmidrule(lr){4-7} \cmidrule(lr){8-10}
& & \textbf{Pairs in Test Set} & \textbf{Seshat} & \textbf{MutationBERT} & \textbf{SODA} & \textbf{WITNESS} & \textbf{Seshat} & \textbf{MutationBERT} & \textbf{SODA} \\
\midrule
\multirow{6}{*}{Chart} & 1 & 101,643& 119.76 & 2,683 & 1,539 & 1.38 & 87$\times$ & 1,944$\times$ & 1,115$\times$ \\
& 5 & 80,626 & 52.93 & 2,150 & 1,229 & 1.22 & 43$\times$ & 1,762$\times$ & 1,007$\times$ \\
& 10 & 74,307 & 48.43 & 1,989 & 1,126 & 1.10 & 44$\times$ & 1,808$\times$ & 1,024$\times$ \\
& 15 & 69,490 & 46.01 & 1,860 & 1,053 & 1.20 & 38$\times$ & 1,550$\times$ & 878$\times$ \\
& 20 & 64,164 & 42.42 & 1,715 & 972 & 0.95 & 45$\times$ & 1,805$\times$ & 1,023$\times$ \\
& 25 & 64,304 & 42.25 & 1,720 & 1,001 & 0.78 & 54$\times$ & 2,205$\times$ & 1,283$\times$ \\
\cmidrule(lr){1-10}
\multirow{6}{*}{JacksonCore} & 25 & 57,348 & 63.97 & 1,500 & 867 & 0.91 & 70$\times$ & 1,648$\times$ & 953$\times$ \\
& 20 & 40,665 & 28.36 & 1,060 & 612 & 0.76 & 37$\times$ & 1,395$\times$ & 805$\times$ \\
& 15 & 32,202 & 23.37 & 839 & 485 & 0.87 & 27$\times$ & 964$\times$ & 557$\times$ \\
& 10 & 33,312 & 23.94 & 868 & 502 & 0.83 & 29$\times$ & 1,046$\times$ & 605$\times$ \\
& 5 & 20,000 & 16.98 & 521 & 300 & 0.57 & 30$\times$ & 914$\times$ & 526$\times$ \\
& 1 & 15,063 & 14.51 & 393 & 226 & 0.52 & 28$\times$ & 756$\times$ & 435$\times$ \\
\cmidrule(lr){1-10}
\multirow{4}{*}{Gson} & 15 & 38,240 & 38.96 & 992 & 575 & 0.67 & 58$\times$ & 1,481$\times$ & 858$\times$ \\
& 10 & 38,064 & 24.04 & 990 & 574 & 0.53 & 45$\times$ & 1,868$\times$ & 1,083$\times$ \\
& 5 & 37,743 & 23.86 & 981 & 568 & 0.54 & 44$\times$ & 1,817$\times$ & 1,052$\times$ \\
& 1 & 23,850 & 17.22 & 619 & 358 & 0.50 & 34$\times$ & 1,238$\times$ & 716$\times$ \\
\cmidrule(lr){1-10}
\multirow{7}{*}{Lang} & 1 & 18,759 & 28.29 & 491 & 282 & 0.64 & 44$\times$ & 767$\times$ & 441$\times$ \\
& 10 & 18,086 & 17.23 & 473 & 272 & 0.58 & 30$\times$ & 816$\times$ & 469$\times$ \\
& 20 & 14,747 & 15.20 & 385 & 221 & 0.55 & 28$\times$ & 700$\times$ & 402$\times$ \\
& 30 & 13,956 & 14.97 & 364 & 210 & 0.54 & 28$\times$ & 674$\times$ & 389$\times$ \\
& 40 & 14,788 & 15.26 & 386 & 222 & 0.54 & 28$\times$ & 715$\times$ & 411$\times$ \\
& 50 & 15,230 & 15.52 & 396 & 228 & 0.60 & 26$\times$ & 660$\times$ & 380$\times$ \\
& 60 & 12,666 & 14.03 & 330 & 190 & 0.53 & 26$\times$ & 623$\times$ & 358$\times$ \\
\cmidrule(lr){1-10}
\multirow{4}{*}{Cli} & 30 & 4,844 & 9.84 & 125 & 72 & 0.42 & 23$\times$ & 298$\times$ & 171$\times$ \\
& 20 & 2.227 & 8.14 & 57 & 33 & 0.41 & 20$\times$ & 139$\times$ & 80$\times$ \\
& 10 & 2,485 & 8.19 & 64 & 37 & 0.42 & 20$\times$ & 152$\times$ & 88$\times$ \\
& 1 & 1,828 & 7.93 & 47 & 27 & 0.42 & 19$\times$ & 112$\times$ & 64$\times$ \\
\cmidrule(lr){1-10}
\multirow{4}{*}{Csv} & 15 & 4,722 & 9.68 & 122 & 70 & 0.42 & 23$\times$ & 290$\times$ & 167$\times$ \\
& 10 & 2,652 & 8.15 & 68 & 39 & 0.41 & 20$\times$ & 166$\times$ & 95$\times$ \\
& 5 & 2,663 & 8.16 & 69 & 39 & 0.41 & 20$\times$ & 168$\times$ & 95$\times$ \\
& 1 & 743 & 8.18 & 19 & 11 & 0.41 & 20$\times$ & 46$\times$ & 27$\times$ \\
\bottomrule
\end{tabular}
    \begin{tablenotes}
	\item Time is measured in seconds. 
    \end{tablenotes}
\end{threeparttable}
}
\label{tab:efficiency_same_version}
\end{table*}

\begin{table*}[htbp]
\centering
\scriptsize
\caption{Time Taken by Major and the Predictive Approaches Along With the Speedup Achieved by WITNESS under the Cross-Version and Cross-Project Scenarios}
\adjustbox{width=1.0\columnwidth}{%
\begin{threeparttable}
\begin{tabular}{@{}cccccccccccc@{}} 
\toprule
 \multirow{3.5}{*}{\textbf{Project Version}} & \textbf{\# Mutant-Test} & \multirow{3.5}{*}{\textbf{Major}} & \multicolumn{4}{c}{\textbf{Prediction Approaches}} & \multicolumn{3}{c}{\textbf{WITNESS Speed Up}} & \\
\cmidrule(lr){4-7} \cmidrule(lr){8-10}
 &  \multirow{-1.5}{*}{\textbf{Pairs}} & & \multirow{2}{*}{\textbf{Seshat}} & \textbf{Mutation-} & \multirow{2}{*}{\textbf{SODA}} & \multirow{2}{*}{\textbf{WITNESS}} & \multirow{2}{*}{\textbf{Seshat}} & \textbf{Mutation-} & \multirow{2}{*}{\textbf{SODA}} & \\
 & \textbf{in Test Set} & & & \textbf{BERT} & & & & \textbf{BERT} & & \\
\midrule
Chart: 25 $\rightarrow$ 1\hphantom{0} & 966,273 & 64,719 & 1,393.43 & 26,085 & 14,764 & 12.02 & 116$\times$ & 2,170$\times$ & 1,228$\times$  & \\
Chart: 25 $\rightarrow$ 5\hphantom{0} & 839,446 & 53,986 & 1,220.24 & 22,630 & 12,773 & 10.61 & 115$\times$ & 2,133$\times$ & 1,204$\times$ & \\
Chart: 25 $\rightarrow$ 10 & 736,020 & 46,983 & 1,032.32 & 19,800 & 11,227 & 9.49 & 109$\times$ & 2,086$\times$ & 1,183$\times$ & \\
Chart: 25 $\rightarrow$ 15 & 727,878 & 46,429 & 1,014.58 & 19,598 & 11,153 & 9.52 & 107$\times$ & 2,059$\times$ & 1,172$\times$ & \\
Chart: 25 $\rightarrow$ 20 & 660,518 & 42,475 & 911.64 & 17,838 & 10,141 & 8.97 & 102$\times$ & 1,989$\times$ & 1,131$\times$ & \\
\cmidrule(lr){1-10}
JC: 1 $\rightarrow$ 25 & 528,887 & 113,343 & 632.60 & 14,109 & 8,052 & 3.93 & 161$\times$ & 3,590$\times$ & 2,049$\times$ & \\
JC: 1 $\rightarrow$ 20 & 392,026 & 88,075 & 432.12 & 10,428 & 5,970 & 3.04 & 142$\times$ & 3,430$\times$ & 1,964$\times$ & \\
JC: 1 $\rightarrow$ 15 & 314,306 & 45,069 & 334.27 & 8,354 & 4,787 & 2.62 & 128$\times$ & 3,189$\times$ & 1,827$\times$ & \\
JC: 1 $\rightarrow$ 10 & 308,935 & 44,110 & 330.33 & 8,212 & 4,700 & 2.56 & 129$\times$ & 3,208$\times$ & 1,836$\times$ & \\
JC: 1 $\rightarrow$ 5\hphantom{0} & 206,189 & 31,257 & 212.90 & 5,491 & 3,145 & 2.04 & 104$\times$ & 2,692$\times$ & 1,542$\times$ & \\
\cmidrule(lr){1-10}
Gson: 1 $\rightarrow$ 15 & 380,421 & 16,738 & 318.77 & 10,265 & 5,789 & 2.76 & 115$\times$ & 3,719$\times$ & 2,097$\times$ & \\
Gson: 1 $\rightarrow$ 10 & 365,494 & 15,986 & 304.76 & 9,812 & 5,563 & 2.77 & 110$\times$ & 3,542$\times$ & 2,008$\times$ & \\
Gson: 1 $\rightarrow$ 5\hphantom{0} & 358,491 & 15,253 & 300.35 & 9,683 & 5,447 & 2.63 & 114$\times$ & 3,682$\times$ & 2,071$\times$ & Cross- \\
\cmidrule(lr){1-10}
Lang: 60 $\rightarrow$ 1\hphantom{0} & 183,722 & 12,924 & 245.56 & 4,885 & 2,787 & 2.50 & 98$\times$ & 1,954$\times$ & 1,115$\times$ & Version \\
Lang: 60 $\rightarrow$ 10 & 176,323 & 13,185 & 220.24 & 4,693 & 2,677 & 2.45 & 89$\times$ & 1,916$\times$ & 1,093$\times$ & \\
Lang: 60 $\rightarrow$ 20 & 144,052 & 5,395 & 182.76 & 3,826 & 2,184 & 2.17 & 84$\times$ & 1,763$\times$ & 1,006$\times$ & \\
Lang: 60 $\rightarrow$ 30 & 143,659 & 5,220 & 179.76 & 3,813 & 2,181 & 2.08 & 86$\times$ & 1,833$\times$ & 1,049$\times$ & \\
Lang: 60 $\rightarrow$ 40 & 143,240 & 4,756 & 176.88 & 3,802 & 2,171 & 1.90 & 93$\times$ & 2,001$\times$ & 1,143$\times$ & \\
Lang: 60 $\rightarrow$ 50 & 150,559 & 6,793 & 184.67 & 3,982 & 2,282 & 2.18 & 85$\times$ & 1,827$\times$ & 1,047$\times$ & \\
\cmidrule(lr){1-10}
Cli: 1 $\rightarrow$ 30 & 58,966 & 1,290 & 52.00 & \textit{1,544} & 890 & 0.67 & 78$\times$ & 2,304$\times$ & 1,328$\times$ & \\
Cli: 1 $\rightarrow$ 20 & 26,540 & 498 & 26.83 & \textit{690} & 399 & 0.58 & 46$\times$ & 1,190$\times$ & 688$\times$ & \\
Cli: 1 $\rightarrow$ 10 & 22,062 & 408 & 23.35 & \textit{574} & 332 & 0.58 & 40$\times$ & 990$\times$ & 572$\times$ & \\
\cmidrule(lr){1-10}
Csv: 1 $\rightarrow$ 15 & 45,343 & 5,289 & 41.57 & 1,191 & 682 & 0.80 & 52$\times$ & 1,489$\times$ & 853$\times$ & \\
Csv: 1 $\rightarrow$ 10 & 24,923 & 1,317 & 25.75 & 659 & 373 & 0.54 & 48$\times$ & 1,220$\times$ & 691$\times$ & \\
Csv: 1 $\rightarrow$ 5\hphantom{0} & 22,968 & 1,179 &24.30 & 607 & 344 & 0.55 & 44$\times$ & 1,104$\times$ & 625$\times$ & \\
\midrule
* $\rightarrow$ JC & 528,887 & 113,343 & 552.73 & 13,882 & 8,368 & 3.23 & 171$\times$ & 4,298$\times$ & 2,591$\times$ & \\
* $\rightarrow$ Gson & 380,421 & 16,738 & 231.53 & 10,473 & 6,010 & 5.09 & 45$\times$ & 2,058$\times$ & 1,180$\times$ & \\
* $\rightarrow$ Lang & 183,722 & 12,924 & 172.97 & 4,833 & 2,884 & 2.32 & 75$\times$ & 2,083$\times$ & 1,243$\times$ &  \\
* $\rightarrow$ Cli\hphantom{0 } & 58,966 & 1,290 & 38.51 & 1,540 & 911 & 0.87 & 44$\times$ & 1,770$\times$ & 1,047$\times$ & Cross- \\
* $\rightarrow$ Csv\hphantom{0} & 45,343 & 5,289 & 31.24 & 1,216 & 682 & 1.05  & 30$\times$ & 1,158$\times$ & 650$\times$ & Project \\
\cmidrule(lr){1-10}
+ $\rightarrow$ JC & 528,887 & 16,738 & 572.50 & 14,281 & 8,026 & 3.35 & 171$\times$ & 4,263$\times$ & 2,396$\times$ & \\
+ $\rightarrow$ Gson & 380,421 & 1,290 & 235.25 & 10,463 & 5,776 & 2.30 & 102$\times$ & 4,549$\times$ & 2,511$\times$ & \\
+ $\rightarrow$ Lang & 183,722 & 5,289 & 191.34 & 4,849 & 2,780 & 1.96 & 98$\times$ & 2,474$\times$ & 1,418$\times$ & \\
\bottomrule
\end{tabular}
    \begin{tablenotes}
         \item JC represents the JacksonCore project.
	\item Time is measured in seconds. 
	\item The asterisk (*) indicates that for the five projects—JacksonCore, Gson, Lang, Cli, and Csv—excluding the project used as the test set, the remaining four are used as the training and validation sets for training the predictive models. The plus sign (+) indicates that Chart, Cli, and Csv are used as the training and validation sets.
    \end{tablenotes}
\end{threeparttable}
}
\label{tab:efficiency}
\end{table*}

\textbf{\textit{Results.}} Table \ref{tab:efficiency_same_version} and Table \ref{tab:efficiency} present the results of efficiency, specifically detailing the time taken for each predictive approach. All times in the two tables are measured in seconds. In Table \ref{tab:efficiency_same_version}, the fourth to seventh columns present the prediction times of the four predictive approaches for predicting the kill matrix on the test set. The last three columns show the speed-up of WITNESS over the three baselines. We do not include the time taken by Major since the test set consists of only 10\% of all mutants in the same-version scenario. In Table \ref{tab:efficiency}, the third column presents the time Major takes to perform actual mutation testing. The fourth to seventh columns present the prediction times of the four predictive approaches for predicting the kill matrix on the test set. The eighth to tenth columns show the speed-up of WITNESS over the three baselines.

As shown in Table \ref{tab:efficiency_same_version} and Table \ref{tab:efficiency}, WITNESS achieves significantly higher efficiency compared to the baselines. The maximum prediction time for WITNESS is 12.02 seconds, recorded when using Chart\_25 to predict Chart\_1. In comparison, the corresponding times consumed by Seshat, MutationBERT, and SODA are 1,393.43, 26,085, and 14,764 seconds, respectively, all of which are significantly longer than the time taken by WITNESS. On average, WITNESS delivers a speed-up in prediction time that is 65.92 times faster than Seshat, 1,722.81 times faster than MutationBERT, and 986.17 times faster than SODA across all scenarios.

The execution time of Seshat that we reran is similar to the results reported in their paper. However, not splitting the test set results in longer prediction times for MutationBERT \cite{jain2023contextual}. Furthermore, MutationBERT reports the time for mutant killing prediction, not for kill matrix prediction. The time for mutant killing prediction is generally faster than that for kill matrix prediction. This is because only one result needs to be predicted per mutant in mutant killing prediction, whereas all results for the formed mutant-test pairs must be predicted when generating the kill matrix for a mutant. These factors help explain why the prediction times we measured for MutationBERT differ significantly from those reported in their paper. In addition, it is important to note that MutationBERT takes longer than Major in the Cli project, as highlighted in italic font in Table \ref{tab:efficiency}. Moreover, although SODA directly modifies the code from MutationBERT, SODA achieves higher efficiency than MutationBERT. This may be because SODA uses smaller checkpoint files during training compared to MutationBERT.

\begin{tcolorbox}[colback=black!5!white, colframe=black!75!black]
\textbf{Answer for RQ5: WITNESS significantly improves efficiency in predicting the kill matrix compared to baselines. On average, WITNESS speeds up the prediction by 65.92, 1,722.81, and 986.17 times compared to Seshat, MutationBERT, and SODA, respectively.}  
\end{tcolorbox}

\subsection{RQ6: Prioritization of Test Cases}

\textbf{\textit{Approach.}} In Defects4J, for each project version, there are two versions: the buggy version and the fixed version. The difference between the two versions lies in the defect. Test case prioritization is based on the predicted kill matrix obtained from the fixed version. This results in a re-ordered test suite, where test cases that kill more mutants are ranked higher. For the prioritized test suite, we perform evaluation based on the buggy version, which includes the real defect. The APFD value is calculated based on the re-ordered test suite and the test cases that can detect the real defect.

We perform test case prioritization based on the predicted kill matrix from different predictive approaches. The prioritized test suite solely involves test cases from mutant-test pairs where mutations occur within source methods. The APFD values computed by Seshat, MutationBERT, SODA, and WITNESS are denoted as $\text{APFD}_{\text{Seshat}}$, $\text{APFD}_{\text{MutationBERT}}$, $\text{APFD}_{\text{SODA}}$, and $\text{APFD}_{\text{WITNESS}}$, respectively. 

In Defects4J, each defect is isolated and lies in only a few Java source classes, with no other real defects present, even though many Java source classes exist. The Defects4J setup results in many Java source classes that are irrelevant to the single defect. For example, in the \textit{source}/ directory of Chart\_1, there are 654 Java source classes, but only one class is modified.

In this paper, mutation testing is performed on all Java source classes of a project version. Having many mutants generated from source classes that are irrelevant to the defect could result in the actual APFD value—calculated using test cases prioritized based on the actual kill matrix—being relatively low. In other words, the test cases prioritized based on the actual kill matrix may not detect real faults earlier.

Additionally, we find that when a test suite is prioritized using a predicted kill matrix with higher predictive performance, the calculated APFD may not increase. Instead, the calculated APFD tends to be close to the actual APFD value. Therefore, we compare these predicted APFD values with the actual APFD values computed from the kill matrix generated by Major, denoted as $\text{APFD}_{\text{ground-truth}}$, which represents the ground truth for the predicted APFD values. The absolute differences between the predicted APFDs and the ground-truth APFD values are represented as $|\text{APFD}_{\text{Seshat}}-\text{APFD}_{\text{ground-truth}}|$, $|\text{APFD}_{\text{MutationBERT}}-\text{APFD}_{\text{ground-truth}}|$, $|\text{APFD}_{\text{SODA}}-\text{APFD}_{\text{ground-truth}}|$, and $|\text{APFD}_{\text{WITNESS}}-\text{APFD}_{\text{ground-truth}}|$. A lower value indicates that the predicted kill matrix is more similar to the actual kill matrix. 

We repeat the prioritization process 10 times for each experiment because if several test cases kill the same number of mutants, the prioritization algorithm randomly selects one of the test cases. Since we repeat the prioritization of each test suite 10 times to reduce randomness, we calculate the average of the absolute differences between the predicted and actual APFD values from these 10 runs.

The time complexity of mutation-based test case prioritization using the additional approach is $O(T^2 M)$, where $T$ is the number of test cases and $M$ is the number of mutants. Specifically, selecting a test case in each iteration involves checking every unselected test case to determine how many mutants it could kill that have not been killed by any previously selected test case, which costs $O(T \cdot M)$. Since there are $T$ test cases, the process is repeated $O(T)$ times. The time complexity of $O(T^2 M)$ means that as the number of test cases increases, the time complexity grows quadratically, which can become computationally expensive for large test suites.

We conduct test case prioritization under same-version, cross-version, and cross-project scenarios. Given that test case prioritization is time-consuming, in the cross-version scenario, we focus on the nearest and farthest two versions. The nearest two versions are adjacent to each other, while the farthest two versions involve using the earliest version to predict subsequent versions. For instance, the nearest versions include Csv\_1 predicting Csv\_5, Csv\_5 predicting Csv\_10, and Csv\_10 predicting Csv\_15. Conversely, the farthest versions involve Csv\_1 predicting Csv\_5, Csv\_10, and Csv\_15.

The projects Chart and Lang have large test suites. For test cases involved in mutant-test pairs where mutations occur inside source methods, the maximum test suite size for the Chart project is 2,176, and for the Lang project, it is 2,198. In one experiment, test case prioritization is performed on four predicted kill matrices generated by the four different predictive approaches. For each predicted kill matrix, we repeat the prioritization process 10 times to reduce randomness, resulting in 40 runs per experiment. There are many different experiments conducted for the two projects, leading to a large total number of runs. Therefore, when using the Additional approach for test case prioritization, we exclude the Lang and Chart projects due to the significantly longer time required to prioritize the entire test suite for experiments in these two projects.

\begin{table*}[hbp]
\centering
\scriptsize
\caption{Mean Absolute Differences Between the Predicted APFDs and the Actual APFDs under the Same-Version Scenario}
\begin{tabular}{ccccccccccc}
\toprule
\multirow{2.5}{*}{\textbf{Project ID}} & \multicolumn{4}{c}{\textbf{Total}} & \multicolumn{4}{c}{\textbf{Additional}} \\
\cmidrule(lr){2-5} \cmidrule(lr){6-9}
& \textbf{Seshat} & \textbf{MutationBERT} & \textbf{SODA} & \textbf{WITNESS} & \textbf{Seshat} & \textbf{MutationBERT} & \textbf{SODA} & \textbf{WITNESS} \\
\midrule
Chart & 0.0302 & 0.1092 & 0.0670 & 0.0452 & - & - & - & - \\
JacksonCore & 0.0394 & 0.0744 & 0.0807 & 0.0363 & 0.1264 & 0.1209 & 0.0764 & 0.1353 \\
Gson & 0.0461 & 0.2820 & 0.0313 & 0.0187 &  0.2200 & 0.3412 & 0.1512 & 0.0460  \\
Lang & 0.1438 & 0.1655 & 0.1166 & 0.1592 & - & - & - & - \\
Cli & 0.0684 & 0.0556 & 0.0697 & 0.0912 & 0.0769 & 0.0725 & 0.0671 & 0.1138 \\
Csv & 0.0790 & 0.0757 & 0.1131 & 0.0473 & 0.1815 & 0.2254 & 0.1745 & 0.1270 \\
\cmidrule(lr){1-9} 
\textit{Avg.} & 0.0678 & 0.1271 & 0.0797 & \textbf{0.0663} & 0.1512 & 0.1900 & 0.1173 & \textbf{0.1055} \\
\bottomrule
\end{tabular}
\label{tab:sapient_same_version}
\end{table*}

\begin{table*}[htp]
\centering
\scriptsize
\caption{Mean Absolute Differences Between the Predicted APFDs and the Actual APFDs under the Cross-Version Scenario}
\begin{tabular}{cccccccccc}
\toprule
\multirow{2.5}{*}{\textbf{Project ID}} & \multicolumn{2}{c}{\textbf{Seshat}} & \multicolumn{2}{c}{\textbf{MutationBERT}} & \multicolumn{2}{c}{\textbf{SODA}} & \multicolumn{2}{c}{\textbf{WITNESS}} & \\
\cmidrule(lr){2-3} \cmidrule(lr){4-5} \cmidrule(lr){6-7} \cmidrule(lr){8-9}
 & \textbf{Nearest} & \textbf{Farthest} & \textbf{Nearest} & \textbf{Farthest} & \textbf{Nearest} & \textbf{Farthest} & \textbf{Nearest} & \textbf{Farthest} & \textbf{Approach} \\
\midrule
Chart & 0.1043 & 0.0802 & 0.0740 & 0.0811 & 0.0763 & 0.0839 & 0.0737 & 0.0604 & \\
JacksonCore & 0.0564 & 0.0646 & 0.0351 & 0.0472 & 0.0494 & 0.0539 & 0.0348 & 0.0658 & \\
Gson & 0.1249 & 0.1345 & 0.3576 & 0.3830  & 0.0906 & 0.1096 & 0.0729 & 0.0913 & \\
Lang & 0.1814 & 0.1617 & 0.1371 & 0.1136 & 0.1435 & 0.1281 & 0.1179 & 0.1081 & Total\\
Cli & 0.0366 & 0.0365 & 0.0778 & 0.0701 & 0.0616 & 0.0350 & 0.0494 & 0.0434 & \\
Csv & 0.0184 & 0.0399 & 0.0467 & 0.0580 & 0.0340 & 0.0380 & 0.0161 & 0.0671 & \\
\cmidrule(lr){1-9} 
\textit{Avg.} & 0.0870 & 0.0862 & 0.1214 & 0.1255 & 0.0759 & 0.0748 & \textbf{0.0608} & \textbf{0.0727} & \\
\midrule
JacksonCore & 0.1706 & 0.2031 & 0.1792 & 0.1306 & 0.1630 & 0.1056 & 0.1315 & 0.1295 & \\
Gson & 0.0806 & 0.0957 & 0.4132 & 0.4343 & 0.1192 & 0.0920 & 0.1071 & 0.0554 & \\
Cli & 0.1092 & 0.1940 & 0.1032 & 0.1207 & 0.0754 & 0.0680 & 0.1388 & 0.1419 & Additional \\
Csv & 0.2193 & 0.2842 & 0.3799 & 0.2482 & 0.2772 & 0.2394 & 0.1910 & 0.2534 & \\
\cmidrule(lr){1-9} 
\textit{Avg.} & 0.1449 & 0.1943 & 0.2689 & 0.2335 & 0.1587 & \textbf{0.1263} & \textbf{0.1421} & 0.1451 & \\
\bottomrule
\end{tabular}
\label{tab:sapient_cross_version}
\end{table*}

\begin{table*}[htp]
\centering
\scriptsize
\caption{Mean Absolute Differences Between the Predicted APFDs and the Actual APFDs under the Cross-Project Scenario}
\begin{tabular}{ccccccccccc}
\toprule
\multirow{2.5}{*}{\textbf{Prediction}} & \multicolumn{4}{c}{\textbf{Total}} & \multicolumn{4}{c}{\textbf{Additional}} \\
\cmidrule(lr){2-5} \cmidrule(lr){6-9}
& \textbf{Seshat} & \textbf{MutationBERT} & \textbf{SODA} & \textbf{WITNESS} & \textbf{Seshat} & \textbf{MutationBERT} & \textbf{SODA} & \textbf{WITNESS} \\
\midrule
One-to-one & 0.0732 & 0.1012 & 0.0493 & 0.0616  & 0.0939 & 0.1194 & 0.1433 & 0.0849 \\
Many-to-one & 0.0727 & 0.1144 & 0.0583 & 0.0457 & 0.1105 & 0.1688 & 0.1827 & 0.1387 \\
\cmidrule(lr){1-9} 
\textit{Avg.} & 0.0730 & 0.1078 & 0.0538 & \textbf{0.0537} & \textbf{0.1022} & 0.1441 & 0.1630 & 0.1118 \\
\bottomrule
\end{tabular}
\label{tab:sapient_cross_project}
\end{table*}

\textbf{\textit{Results.}} Table \ref{tab:sapient_same_version}, Table \ref{tab:sapient_cross_version}, and Table \ref{tab:sapient_cross_project} show the mean absolute differences between the predicted APFDs and the actual APFDs obtained from the four predictive approaches using the Total and Additional approaches under different scenarios. In the \textit{Avg.} row of the three tables, superior values are highlighted in bold. 

As shown in the three tables, the absolute differences achieved by different predictive approaches vary across projects, scenarios, and prioritization approaches. On average, WITNESS generally achieves the lowest absolute differences in APFD values compared to the baselines across different scenarios. Exceptions occur under the cross-version scenario using the Additional approach with the farthest prediction, where SODA achieves the lowest average absolute differences. This is due to the underperformance of WITNESS in the Cli project, attributed to its lower predictive performance compared to SODA in determining whether a mutant will be killed. Similarly, under the cross-project scenario using the Additional approach, Seshat records the lowest average absolute differences. Specifically, WITNESS outperforms Seshat in the one-to-one prediction but underperforms in the many-to-one prediction.

\begin{tcolorbox}[colback=black!5!white, colframe=black!75!black]
\textbf{Answer for RQ6: WITNESS generally achieves APFD values that are close to the actual APFD values when compared to the baseline approaches. This indicates that the kill matrix predicted by WITNESS is more similar to the actual kill matrix.} 
\end{tcolorbox}

\section{Discussion}

In this section, we present details of WITNESS, including its features, models, and optimal thresholds. Additionally, we discuss the implications for researchers and practitioners, as well as the threats to the validity of our study.

\subsection{Feasibility of extending to other programming languages} The feature set of WITNESS is not limited to statically typed languages. Although the feature \textit{DeclaredVariableType} captures the types of declared variables, its values are abstracted into seven main categories. Thus, it is feasible to extend WITNESS to dynamically typed languages. In dynamically typed programming languages such as Python, there is no need to explicitly declare the type of a variable. However, every object in Python has a type, which is determined at runtime. When a value is assigned to a variable, Python internally creates an object in memory, and this object has a specific type. Specifically, the built-in \textit{type()} function is used to determine the data type of an object. For example, the output of \textit{type(3.14)} is $<$class ‘float’$>$. Moreover, while features that are not constrained to a specific programming language may offer better generality, features tailored to a specific language could provide potential benefits for improving predictive performance within that language.

We perform project analysis to collect features for WITNESS. This project analysis overcomes the limitations of MutationBERT and SODA, which fine-tune CodeBERT and CodeT5, both of which have input limits of 512 tokens. MutationBERT and SODA require specific processing to comply with these token limits. For MutationBERT and SODA, the token limitation constrains the contextual information available for predicting the kill matrix.

Since WITNESS relies on program analysis to extract features, the main challenge in extending our approach to other languages lies in the need for accurate program analysis. This analysis requires parsing the source code, and because grammar varies across programming languages, source code parsing is language-specific.

\subsection{Optimal Threshold}

In predicting the kill matrix, the mutant-test pairs where the test case kills the mutant are typically the minority class, which introduces a class imbalance problem in the classification task. Many binary classification tasks default to using 0.5 as the threshold. However, when the positive class is the minority, the default threshold of 0.5 may result in many actual positive cases being misclassified as negative. Therefore, it is common to adjust the decision threshold to enhance prediction performance for the minority class. 

There are noticeable differences between the thresholds calculated from the Precision-Recall curve and the ROC curve. Since the results are similar for all mutant-test pairs and for those where mutations occur within source methods, we focus our analysis specifically on the latter.

\begin{figure}[htbp]
    \centering
        \includegraphics[scale=0.20]{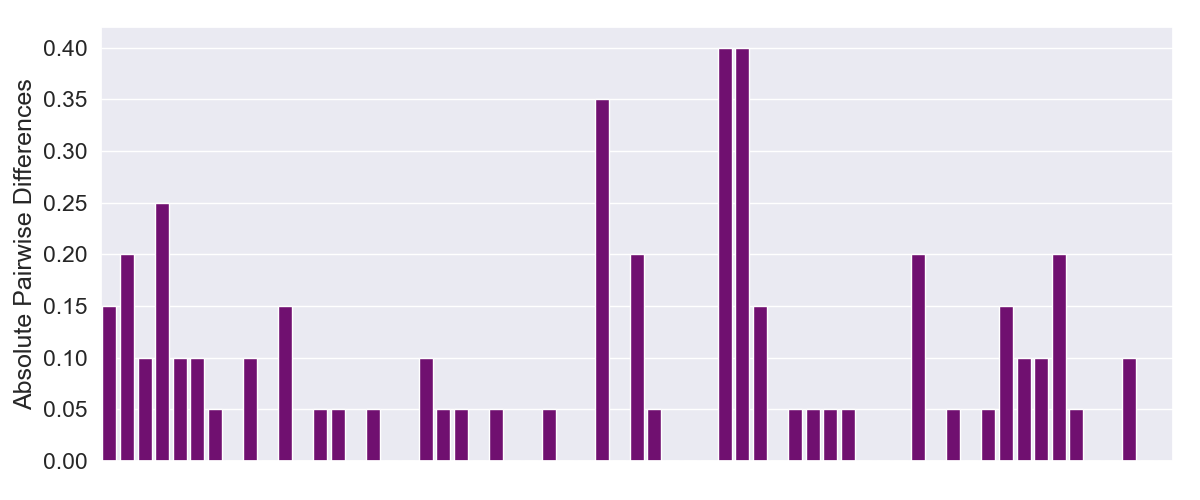}
    \caption{Absolute differences between the thresholds calculated from the Precision-Recall curve and the ROC curve in the validation sets for mutant-test pairs where mutations occur within source methods across all scenarios.}
    \label{fig:pr_roc}
\end{figure}

Fig. \ref{fig:pr_roc} presents the absolute differences between these thresholds in the validation sets across all scenarios. Since our evaluation includes 128 experiments, we streamline the presentation by selecting only one experiment from those with identical training and validation sets, as the calculated thresholds remain the same across these experiments. In Fig. \ref{fig:pr_roc}, the x-axis represents all the selected experiments. As shown in Fig. \ref{fig:pr_roc}, while some thresholds calculated from the Precision-Recall curve and the ROC curve may be the same, a non-negligible proportion of them differ. Notably, the maximum observed difference is 0.40, which occurs when using Cli\_10 and Cli\_20 as the training set under the cross-version scenario.

Thus, combining the F1-score from the Precision-Recall curve and the Youden’s J statistic from the ROC curve could help derive a more balanced threshold that takes into account both precision-recall trade-offs and true/false positive rates. For example, under the same-version scenario, the APE obtained using the threshold computed solely from the Precision-Recall curve is 6.92, while the APE based on the ROC curve is 7.69. Using the combined threshold, the APE is reduced to 6.87. Therefore, we adopt a combined threshold derived from both the Precision-Recall curve and the ROC curve as the optimal threshold.

\begin{figure}[htbp]
    \centering
    \subfigure[Same-Version]{
        \includegraphics[scale=0.15]{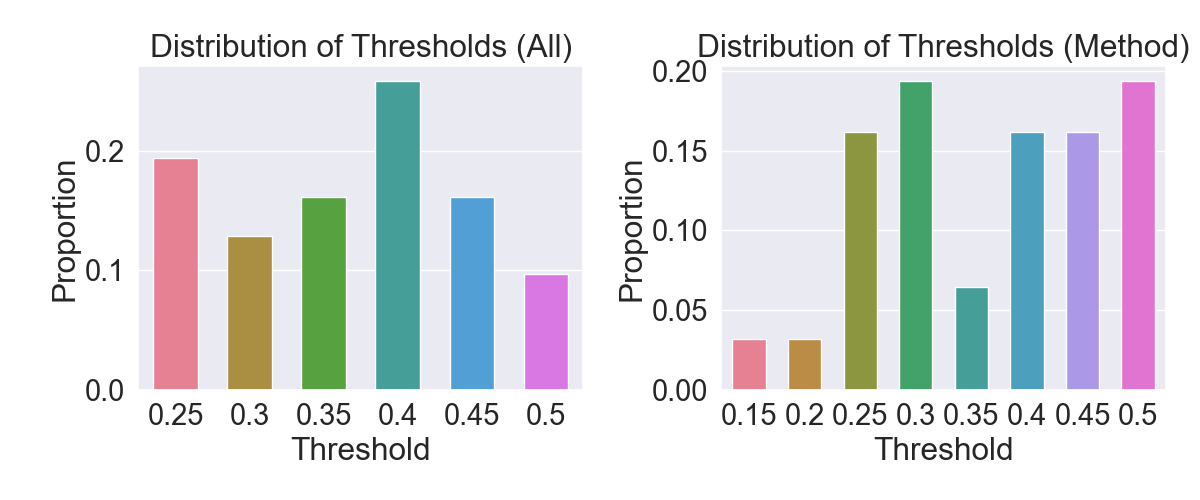}
    }
    \vfill
    \centering
    \subfigure[Cross-Version]{
        \includegraphics[scale=0.15]{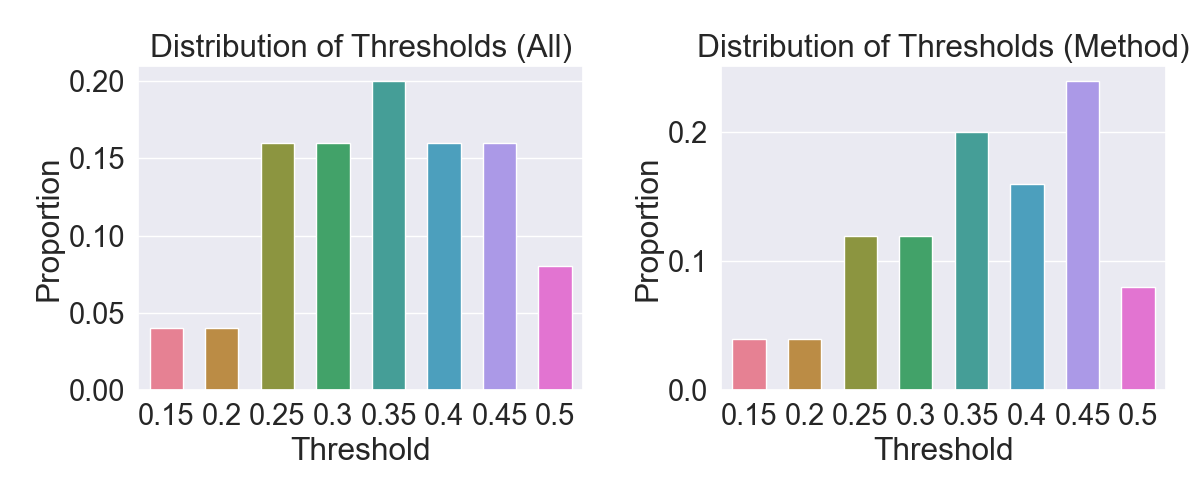}
    }
    \vfill
    \centering
    \subfigure[Cross-Project]{
        \includegraphics[scale=0.15]{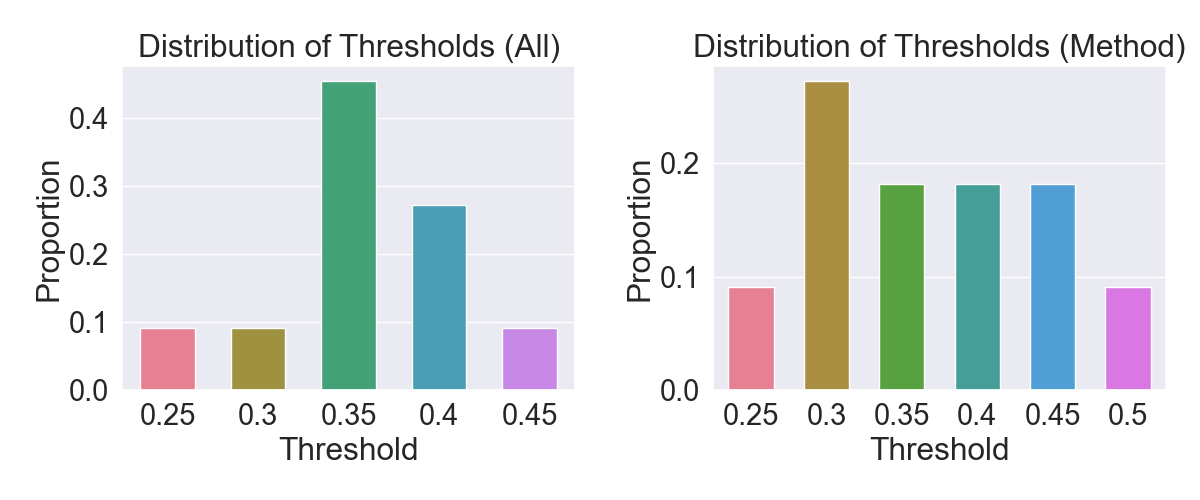}
    }
    \caption{Distribution of optimal thresholds trained on all mutant-test pairs and on those pairs where mutations occur within source methods.}
    \label{fig:optimal_thresholds}
\end{figure}

Fig. \ref{fig:optimal_thresholds}(a), Fig. \ref{fig:optimal_thresholds}(b), and Fig. \ref{fig:optimal_thresholds}(c) display the optimal thresholds for all mutant-test pairs, as well as for those pairs where mutations occur within source methods, across same-version, cross-version, and cross-project scenarios in the validation set, respectively. Notably, the thresholds 0.25, 0.3, 0.35, 0.4, 0.45 are identified as optimal, representing a significant proportion of the experiments for both categories of mutant-test pairs. For the evaluation in the test set across all our experiments, we have chosen 0.35 as the threshold. This choice is based on the fact that 0.35 lies in the middle of the range from 0.25 to 0.45, offering a trade-off among the various optimal thresholds that fall below and above it.

\subsection{Ablation Study}

WITNESS uses combined prediction results from both Random Forest and LightGBM models. We perform an ablation study to demonstrate that the combination can overcome the drawbacks of any single model, thus achieving better prediction performance.

We conduct the ablation study under the same-version scenario and many-to-one prediction under the cross-project scenario, which is sufficient to demonstrate the benefits of combining predictions from the two models. The ablation study is performed on all mutant-test pairs. For the thresholds used to classify predictions from each of the Random Forest and LightGBM models, we adopt 0.35, consistent with the threshold used when combining predictions from both models.

\begin{table*}[htbp]
\centering
\scriptsize
\caption{Mean and Median Predictive Performance of Combined and Separate Predictions}
\begin{tabular}{cccccccc}
\toprule
\multirow{2.5}{*}{\textbf{Scenario}} & \multicolumn{2}{c}{\textbf{Random Forest}} & \multicolumn{2}{c}{\textbf{LightGBM}} & \multicolumn{2}{c}{\textbf{Combined}} & \\
\cmidrule(lr){2-3} \cmidrule(lr){4-5} \cmidrule(lr){6-7}
& \textbf{Mean} & \textbf{Median} & \textbf{Mean} & \textbf{Median} & \textbf{Mean} & \textbf{Median} & \\ 
\midrule
Same-Version & 0.725 & 0.750 & 0.728 & 0.760 & \textbf{0.735} & \textbf{0.770} & Kill Matrix \\
Cross-Project & 0.540 & 0.535 & 0.536 & 0.530 & \textbf{0.543} & \textbf{0.540} & Prediction \\ 
\midrule
Same-Version & \textbf{0.682} & 0.680 & 0.615 & 0.660 & 0.669 & \textbf{0.690} & Mutant Killing \\
Cross-Project & 0.404 & 0.390 & 0.424 & \textbf{0.425} & \textbf{0.428} & 0.410 & Prediction \\ 
\midrule
Same-Version & 5.765 & 4.630 & 6.336 & 4.950 & \textbf{5.583} & \textbf{3.190} & Mutation Score \\
Cross-Project & 8.149 & 8.690 & 6.831 & 6.050 & \textbf{5.471} & \textbf{4.990} & Prediction \\ 
\bottomrule
\end{tabular}
\label{tab:hard_to_kill}
\end{table*}

Table \ref{tab:hard_to_kill} presents the mean and median predictive performance of the combined prediction and the separate predictions from Random Forest and LightGBM, respectively. Kill matrix prediction and mutant killing prediction are measured by F1-score, while mutation score prediction is measured by APE. The best results in Table \ref{tab:hard_to_kill} are shown in bold. While predictions from Random Forest alone or LightGBM alone can sometimes achieve better results than the combined predictions, the combined prediction generally achieves the best overall performance.

\subsection{Implications}

\subsubsection{\textbf{For researchers, it is not necessary to seek out complex models to perform fine-grained predictive mutation testing}}

Seshat was the first approach to predict kill matrices in predictive mutation testing. Its intuition is based on the semantic similarity between the names of source methods and test methods. Seshat learns the relationship between a test case and a mutant using a model that consists of word embedding layers, bidirectional GRU layers, comparison layers, and a final linear layer. Later, MutationBERT fine-tuned CodeBERT \cite{feng2020codebert} to predict the kill matrix. The fine-tuned CodeBERT is a more complex model than the one Seshat adopts. SODA first applies contrastive learning to capture subtle differences in mutants, followed by fine-tuning CodeT5. The model trained by SODA is also more complex than the one trained by Seshat.

In terms of efficiency, training the predictive models and predicting killing results for all baselines requires available GPUs to accelerate the processes. For MutationBERT and SODA, even when both training and prediction are conducted on GPUs, these approaches require significantly more time. Specifically, MutationBERT takes significantly longer than other predictive approaches to produce results on the test set, exceeding the time taken by actual mutation testing in all adopted versions of the Cli project. Since the purpose of predictive mutation testing is to reduce the cost of mutation testing, solely pursuing higher predictive performance at the cost of increased resource and time consumption may undermine the goal of reducing the high cost of mutation testing.

The experimental results of WITNESS demonstrate that, for the specific task of fine-grained predictive mutation testing, simpler classical machine learning approaches can be more effective and efficient. However, we do not deny the significant value of deep learning in software engineering research. Our study suggests that future research should more critically consider how deep learning is applied in this domain. The focus should be on when and how its unique strengths can be harnessed to deliver tangible benefits, rather than adopting it by default.

\subsubsection{\textbf{For practitioners, it provides a practical method to adopt fine-grained predictive mutation testing for applications}}

WITNESS overcomes the limitations of three previous studies that predict the kill matrix solely involving mutant-test pairs with mutations inside source methods. WITNESS is not only suitable for all generated mutants but also enhances the effectiveness and efficiency of fine-grained predictive mutation testing. This improvement reduces the cost of mutation testing and aids in evaluating test suite effectiveness in a lightweight manner, advancing its practical application.

Based on the calculated optimal thresholds across all scenarios, the implication is that selecting a threshold lower than 0.5 is a better choice for achieving higher predictive performance, as the majority of optimal thresholds fall below 0.5. Regarding how to set thresholds for future developers, it is advisable to determine the threshold empirically. However, a threshold of 0.35 is a reasonable choice, as the optimal threshold varies significantly across experiments—even within the same scenario—typically ranging from 0.25 to 0.45.

Given that WITNESS demonstrates superior predictive performance compared to baselines, it also suggests another practical application: test suite augmentation. Generally, the survived mutants highlight weaknesses in a test suite \cite{dakhel2024effective}. Enhancing the effectiveness of test suites can be achieved by generating additional test cases that target and kill the survived mutants. Using WITNESS enables practitioners to lower the costs associated with mutation testing and use the predictions of survived mutants to guide the generation of additional test cases. Therefore, the effectiveness of the test suite can be enhanced by WITNESS in a lightweight manner.

\subsection{Threats to Validity}

We discuss the construct validity, internal validity, and external validity of our study. Construct validity is concerned with whether the study accurately measures or represents the concept it claims to. Interval validity assesses whether the changes in the dependent variable are indeed caused by the manipulation of the independent variable, and not by other factors. External validity refers to the extent to which the results of the study can be generalized to other research settings, populations, and conditions. 

\textbf{\textit{Construct Validity:}} To collect static features, we utilize ANTLR and Understand, both of which are commonly used in existing studies \cite{li2020securing, holler2012fuzzing, zhou2014depth, wang2017qtep}. WITNESS includes one dynamic feature, \textit{HitsNumber}, which requires a code coverage tool to collect. In this paper, we use Cobertura to collect the dynamic feature, a tool widely used in previous studies \cite{zhang2020cbua, zhang2023assessing}. Using these tools mitigates the threat to the accuracy of the measurements. Another threat to construct validity comes from the variation in inference time across different executions. To mitigate this threat, we conduct multiple runs and use the average time from these repetitions. The manually engineered features in WITNESS may not be complete. However, WITNESS generally achieves higher predictive performance than the baselines by combining features from three categories: source code, source code changes, and test cases. We leave the exploration of automated feature learning for future work.

\textbf{\textit{Internal Validity:}} The main threat to internal validity is the predictive models we used. These models play a key role in making predictive decisions. We mitigate this threat by using Random Forest and LightGBM, which are commonly used machine learning models. We have analyzed the importance of various features used in WITNESS, enhancing the understanding of the decision-making processes of the two adopted machine learning models. Another threat to internal validity is the use of a single, fixed dataset split without repetitive splitting. However, repetitive dataset splitting requires extensive computational resources. Our thorough evaluation, encompassing 128 experiments across three scenarios, serves to mitigate this threat.

\textbf{\textit{External Validity:}}  We conducted experiments with projects from Defects4J, which are commonly used in baseline studies. These projects are open-source, well-maintained, and thoroughly tested. The projects, categorized as Java SE and Java EE projects, span a variety of application domains. For instance, the project Chart is a library for producing charts, and the project Cli provides an API for parsing command-line options passed to programs. While our results may not fully generalize to all projects, the comprehensive evaluation involving 128 experiments across diverse scenarios helps mitigate this threat. In this study, for actual mutation testing, we use Major, which is commonly adopted by all baselines. Other mutation testing tools, such as PIT \cite{coles2016pit}, also exist. However, PIT performs mutations at the bytecode level, which may not accurately reflect the corresponding changes in the source code. As future work, we plan to extend our study to include mutants generated by other mutation testing tools.

\section{Conclusions and future work}

In this paper, we propose WITNESS, which employs classical machine learning to predict the kill matrix. Although WITNESS is simpler, it generally achieves higher predictive performance than the baselines. Its high effectiveness benefits test case prioritization conducted on the predicted kill matrix. WITNESS eliminates the high computational costs of GPU-based training and prediction, achieving significantly greater prediction efficiency compared to the baselines. It also overcomes the limitation of baselines that cannot handle all generated mutants, making its practical use feasible. 

Our feature importance analysis in WITNESS provides guidance for future research, highlighting the value of prioritizing features related to source code changes. In the future, we plan to adapt WITNESS for test suite augmentation, targeting predicted survived mutants to enhance test suite effectiveness efficiently.

\section*{Data Availability}

We have released our dataset and source code online \cite{lu2024supplementary}, including detailed instructions for reproducing our experiment.

\section*{Appendix. Details of the features in WITNESS} 
We specifically provide details on several categorical features, whose full set of categories is not discussed in Section 3.2, to facilitate replication of our approach.

\textbf{\textit{StatementType.}} The implementation of WITNESS supports the following Java statement types:

\begin{itemize}
    \item \textit{MemberDeclaration}, \textit{LocalVariableDeclaration}
    \item \textit{EnumConstant}
    \item \textit{ASSERT}, \textit{IF}, \textit{FOR}, \textit{WHILE}, \textit{TRY}, \textit{SWITCH}, \textit{YIELD}
    \item \textit{RETURN}, \textit{THROW}, \textit{BREAK}, \textit{CONTINUE}, \textit{SYNCHRONIZED}
    \item \textit{methodCall, methodCall-super, methodCall-this, methodCall-set, methodCall-get}
    \item \textit{ASSIGN}, \textit{INC}, \textit{DEC}, \textit{AND\_ASSIGN}, \textit{OR\_ASSIGN}, \textit{XOR\_ASSIGN}, \textit{ADD\_ASSIGN}, \textit{SUB\_ASSIGN}, \textit{MUL\_ASSIGN}, \textit{DIV\_ASSIGN}, \textit{MOD\_ASSIGN}, \textit{LSHIFT\_ASSIGN}, \textit{RSHIFT\_ASSIGN}, \textit{URSHIFT\_ASSIGN}
    \item \textit{expression}
\end{itemize}

Although these statement types do not cover all possible Java statements, they are sufficient to distinguish the statements in the six adopted projects. The statement types are based on ANTLR grammar, and our implementation can be easily extended to support additional statement types.

\textbf{\textit{ParentContextType.}} This feature represents the type of the parent context, similar to the feature \textit{StatementType}. Specifically, the unique values of \textit{ParentContextType} include:

\begin{itemize}
\item \textit{ConstructorDeclaration}: if the mutated statement is a child of a constructor in the parse tree.
\item \textit{MethodDeclaration}: if the mutated statement is a child of a source method in the parse tree.
\item \textit{Block}: if the mutated statement is a child of an anonymous block in the parse tree.
\end{itemize}

\textbf{\textit{DeclaredVariableType.}} The value of this feature is abstracted and categorized as \textit{NUMERIC}, \textit{CHAR}, \textit{BOOLEAN}, \textit{STRING}, \textit{COLLECTION}, \textit{MAP}, or \textit{OBJECT}. Specifically, \textit{NUMERIC} is set for concrete primitive data types — byte, short, int, long, float, double — and their wrapper classes. \textit{COLLECTION} is set for classes that extend or implement java.util.Collection. \textit{MAP} is set for classes that extend or implement java.util.Map. \textit{OBJECT} is set if the data type of a declared variable does not belong to any of the other six categories. Additionally, \textit{ARRAY} is appended to any of the seven types if the variable is an array.

\textbf{\textit{SkeletonModification.}} The \textit{SkeletonModification} feature represents changes in conditions at an abstract level, providing a high-level perception of condition mutations. This feature is non-empty for conditional statements, specifically for while or do-while statements, switch statements, return statements, and for loops.

For each condition, we recursively abstract each sub-condition. The sub-conditions are separated based on \&\& and $||$ operators. If the number of sub-conditions differs before and after mutation, we abstract each sub-condition as \textit{expri}. For example, if the condition \textit{allStringsNull $||$ longestStrLen == 0 \&\& !anyStringNull} is mutated to \textit{longestStrLen == 0 \&\& !anyStringNull}, then the \textit{SkeletonModification} feature is a list of two elements: [\textit{expr1 $||$ expr2 \&\& expr3}, \textit{expr1 \&\& expr2}]. If the condition before and after mutation has the same number of sub-conditions, the abstraction of each sub-condition becomes more concrete. In this case, we include relational operators (==, !=, $>$, $<$, $\ge$, $\le$) in the abstracted sub-conditions. For example, if the condition \textit{src.length $>$ srcPos + 1 \&\& src[srcPos + 1]} is mutated to \textit{false \&\& src[srcPos + 1]}, then the \textit{SkeletonModification} feature is [\textit{(expr1 $>$ expr2) \&\& expr3}, \textit{expr1 \&\& expr2}]. This abstraction provides detailed information about mutations that modify sub-conditions, such as changing a greater-than relation to a non-relational sub-condition.

\bibliographystyle{ACM-Reference-Format}
\bibliography{sample-base}

\end{document}